\documentclass[11pt,a4paper]{article}
\usepackage[margin=2.2cm]{geometry}

\usepackage{lineno}   
 
\usepackage[english]{babel}
\usepackage[utf8]{inputenc}
\usepackage{amsmath}
\usepackage{enumitem}
\usepackage{amsthm}
\usepackage{xcolor}

\newtheorem{definition}{Definition}
\newcommand{\cS}{\mathcal{S}}
\usepackage{graphicx}
\usepackage{pifont}
\usepackage{booktabs}
\usepackage{braket}
\usepackage{mathrsfs}
\usepackage{color}
\usepackage{latexsym,amsmath,verbatim}
\usepackage{amssymb} 
\usepackage{amsfonts}
\usepackage{colortbl}
\usepackage{array}
\usepackage{stackrel}
\usepackage{bm}
\usepackage{nicefrac}
\usepackage{rotating}
\usepackage{float}
\usepackage[unicode=true,
bookmarks=true,bookmarksnumbered=false,bookmarksopen=false,
 breaklinks=false,pdfborder={0 0 0},pdfborderstyle={},backref=false,colorlinks=true]
 {hyperref}
 \hypersetup{
 citecolor=blue,
 urlcolor=blue,
 linkcolor=blue}

\usepackage[final]{pdfpages}

\makeatletter
\def\sf@plv{1}
\def\sf@correlation{2}
\def\sf@perolation_cliques{3}
\def\sf@comparison_connectome{4}
\def\sf@comparison_rs{5}
\def\sf@comparison_task{6}
\def\sf@extended_figure_glasser{7}
\def\sf@variationcoefficient{8}
\def\sf@jsd{9}
\def\sf@extended_figure_12{10}
\def\sf@extended_figure_deconvolved{11}
\def\sf@extended_figure_deconvolved_glasser{12}
\def\sf@bandpass{13}
\newcommand{\SuppFig}[1]{Supplementary Fig.~\@nameuse{sf@#1}}
\makeatother

\title{Fibration symmetry-breaking supports functional transitions in
  a brain network engaged in language}
  
\author{\parbox{0.95\textwidth}{\centering Tommaso Gili$^{1,2}$, Bryant Avila$^3$, Luca Pasquini$^{4,5}$, Andrei Holodny$^{4,6,7}$, David Phillips$^{8,9}$, Paolo Boldi$^{10}$, Andrea Gabrielli$^{11,12}$, Guido Caldarelli$^{2,13,14}$, Manuel Zimmer$^{15}$ \& Hern\'an A. Makse$^{3,4,\ast}$}}

\date{}
\begin{document}
\makeatletter\g@addto@macro\@floatboxreset{\def\baselinestretch{1}\small}\makeatother 

\maketitle

\begin{enumerate}[label=\textsuperscript{\arabic*},leftmargin=*,nosep]
 \item Networks Unit, IMT Scuola Alti Studi Lucca, Piazza San Francesco 15, 55100- Lucca, Italy.
 \item Institute for Complex Systems (ISC), CNR, UoS Sapienza, Rome, 00185, Italy.
 \item Levich Institute and Physics Department, City College of New York, New York, NY 10031, USA.
 \item Neuroradiology Service, Department of Radiology, Memorial Sloan Kettering Cancer Center, New York, NY 10065, USA.
 \item Neuroradiology Unit, NESMOS Department, Sant’Andrea Hospital, La Sapienza University, Rome, 00189, Italy.
 \item Department of Neurology and Neuroscience, Weill Medical College of Cornell University, New York, NY, 10021, USA.
 \item Department of Radiology, Weill Medical College of Cornell University, New York, NY 10065, USA.
 \item Division of Mathematics, Computer and Information Systems, Office of Naval Research, Arlington, VA 22217, USA.
 \item Department of Mechanical Engineering,
        University of New Mexico, Albuquerque, NM 87131, USA.
 \item Department of Computer Science,
        University of Milan, Milano, Italy.
 \item Enrico Fermi Research Center (CREF), Via Panisperna 89A, 00184 - Rome, Italy.
 \item Dipartimento di Ingegneria Civile, Informatica e delle Tecnologie Aeronautiche, Universit\`a degli Studi Roma Tre,        Via Vito Volterra 62, 00146 - Rome, Italy.
 \item Department of Molecular Science and Nanosystems and ECLT, Ca' Foscari University of Venice, Venice, 30123, Italy.
 \item London Institute for Mathematical Sciences, Royal Institution, 21 Albemarle St London W1S 4BS, UK.
 \item Research Institute of Molecular Pathology (IMP), Vienna Biocenter (VBC), Campus-Vienna-Biocenter 1, 1030 Vienna,             Austria.
 
\end{enumerate}

\noindent $^{\ast}$Corresponding author. Email: hmakse@ccny.cuny.edu

\clearpage

\begin{abstract}
Symmetry and symmetry breaking describe physical phenomena from particle interactions to phase transitions, but their applicability to biological systems remains limited. Here we study how the structural connectivity of the human brain supports synchronization of mesoscopic regions engaged in language. We describe this relationship through symmetry fibrations, local symmetries that predict cluster synchronization from the input structure of a network. Using functional magnetic resonance imaging data from 20 healthy participants at rest and during two language tasks, we apply a mixed integer linear programming algorithm to infer the minimal subgraph of a literature-curated dorsal-stream connectome consistent with each observed synchronization pattern. The global symmetry of the baseline connectome is lost at rest, where left-right fibration symmetry is retained, and fibration symmetry is further broken in the transition from rest to language, in agreement with left lateralization. These results indicate that local network symmetry constrains how brain structure supports coherent function.
\end{abstract}

\section{Introduction}
\label{sec:intro}

The network of internal connections shapes collective phenomena in complex dynamical systems \cite{barzel2013universality}. In particular, synchronization, a collective behavior in which the dynamics of the network nodes converge on the same time evolution, can be exhibited either as a global state \cite{arenas2008synchronization}, in which all units follow the same trajectory, or through clustered states, in which the system splits into subsets of units synchronized with each other \cite{pecora2014cluster,sorrentino2016complete}. In the latter phenomenon, known as cluster synchronization (CS) \cite{pecora2014cluster,sorrentino2016complete}, the composition of the clusters is determined by the symmetries of the network of connections \cite{pecora2014cluster,sorrentino2016complete,golubitsky2006nonlinear,stewart2003symmetry,boldi2002fibrations,deville2015modular,nijholt2016graph,morone2019symmetry,morone2020fibration,leifer2020circuits}. The symmetries of a network can therefore predict the emergence of CS in a dynamics defined on it.

Relating structure to function is a long-standing problem in systems neuroscience \cite{honey2009, bullmore2009complex, friston2011functional, friederici2011brain, park2013structural, hermundstad2013structural, goni2014resting, chen2017features, deco2014identification, preti2019decoupling}. Earlier empirical studies of the structure-function relationship in the human brain have used diffusion tractography and functional magnetic resonance imaging (fMRI) to correlate white-matter tracts (structural connectivity) with resting-state functional coupling between regions of interest (ROIs) \cite{honey2009,koch2002an,greicius2009resting,vandenheuvel2009functionally}. Although structural connectivity partially resembles resting-state functional connectivity, two ROIs can be structurally connected but not functionally related, and vice versa. More recently, the structure-function relationship has been investigated by neurodynamical modeling of fMRI signals in resting and task-based cognitive states \cite{6deco2013,6wein2021}, validated by comparing predicted spatiotemporal patterns with empirical functional connectivity data, and geometric constraints on curvature and distance have been shown to shape spontaneous and induced brain activity \cite{pang2023geometric}. An extensive literature on the network structure of blood oxygen level-dependent (BOLD) signals across task states \cite{cole2014intrinsic,krienen2014reconfigurable,hearne2017reconfiguration, gratton2018functional, ito2020task, cole2021functional, mastrandrea2023information} has identified the network architectures and reconfiguration signatures that distinguish task states from rest. However, a single structural connectome supports many functional connectivity patterns, so that structure alone does not determine which pattern is expressed in a given state (one-to-many degeneracy) \cite{park2013structural}. The correspondence between the functional connectivity patterns associated with functional states (rest or tasks) and the specific structural architecture that sustains their communication dynamics therefore remains an open question \cite{park2013structural}.

Here we address this question with the theory of symmetry and symmetry breaking \cite{anderson1972more}, widely used in physics \cite{weinberg1995thequantum,georgi2000lie,wilczek2016beautiful}, geometry \cite{weyl1952symmetry}, dynamical systems \cite{golubitsky2006nonlinear,stewart2003symmetry,pecora2014cluster,sorrentino2016complete} and geometric deep learning \cite{bronstein2021geometric}. The symmetries relevant for the brain are not those of physical systems. Physical (and geometric) symmetries are automorphisms and form symmetry groups \cite{georgi2000lie,weinberg1995thequantum}. These are global symmetries, since they preserve the global shape of objects; in the case of graphs, they are permutations of nodes that preserve the global adjacency of nodes. The symmetries we find in the brain network are instead symmetry fibrations \cite{boldi2002fibrations,morone2020fibration}, derived from Grothendieck fibrations in category theory \cite{grothendieck1959technique}, which form symmetry groupoids \cite{golubitsky2006nonlinear,stewart2003symmetry}. Fibrations are less restrictive than automorphisms because they are local symmetries that preserve only the isomorphic inputs of nodes; consequently, they preserve the dynamical evolution that leads to cluster synchronization in the network. Fibrations describe how genetic networks are built from the bottom up to process information through gene expression \cite{golubitsky2006nonlinear,morone2019symmetry,morone2020fibration,leifer2020circuits,leifer2021predicting}, appear in simple neural networks such as that of {\it C. elegans} \cite{morone2019symmetry,avila2025symmetries,avila2024fibration,leifer2022symmetry,boldi2022quasifibrations}, and explain the expressive power of graph neural networks \cite{xu2018powerful,velarde2024uncovering}.

We extend this view to the human brain, letting the empirical synchronization activity drive the inference of the underlying connectome. With a symmetry-driven algorithm based on mixed integer linear programming (Fig.~\ref{fig:intro}), we infer the structural network that sustains the cluster synchronization of the functional language network, the cortical sub-network involved in language \cite{friederici2011brain}, from the BOLD synchronization patterns measured at rest and during two language tasks (Fig.~\ref{fig:functional}). The baseline connectome is a binary bidirected graph that encodes the known white-matter pathways of the dorsal language stream, curated from the neuroanatomical literature, rather than a matrix of subject-specific diffusion tracts. Rather than deriving functional organization from an empirically measured structural connectome, we use the observed CS patterns to infer the minimal subgraph of the baseline connectome that is consistent with each functional state via fibration symmetry, removing the connections not required by that state. Inferred subgraphs represent the minimal structural constraints required to maintain each functional state, not a claim about the complete biological substrate. In this way, the fibration framework addresses the degeneracy of the structural network rather than of the functional network, associating each functional state with a minimal structural subgraph.

In analogy with the theory of phase transitions in physical systems \cite{anderson1972more,yeomans1992statistical}, we describe the recruitment of communication resources across brain states as a process of network symmetry breaking. We find that the baseline connectome of the language network displays a global group symmetry, which switches to a local fibration symmetry to sustain the resting-state synchronization dynamics. This fibration symmetry is then further broken by the activity-driven lateralization induced by the language tasks.

Identifying symmetry at the level of the structural pathways complements previous graph studies of the functional network: knowing the anatomical substrate of mesoscale dynamics is a key step towards understanding the functional network \cite{park2013structural}. The framework may also be used to study neurologic and psychiatric conditions known to affect information transmission in brain networks, such as substance abuse, depression, schizophrenia or Alzheimer's disease \cite{bullmore2009complex}, by identifying the mesoscopic pathways that sustain a given functional state as candidate biomarkers and targets for intervention.

\section{Results}
\label{sec:results}

\subsection{Theory of global and local symmetries}
\label{sec:conn_clust}

\begin{figure*}[t!]
\centering
\includegraphics[width=\textwidth]{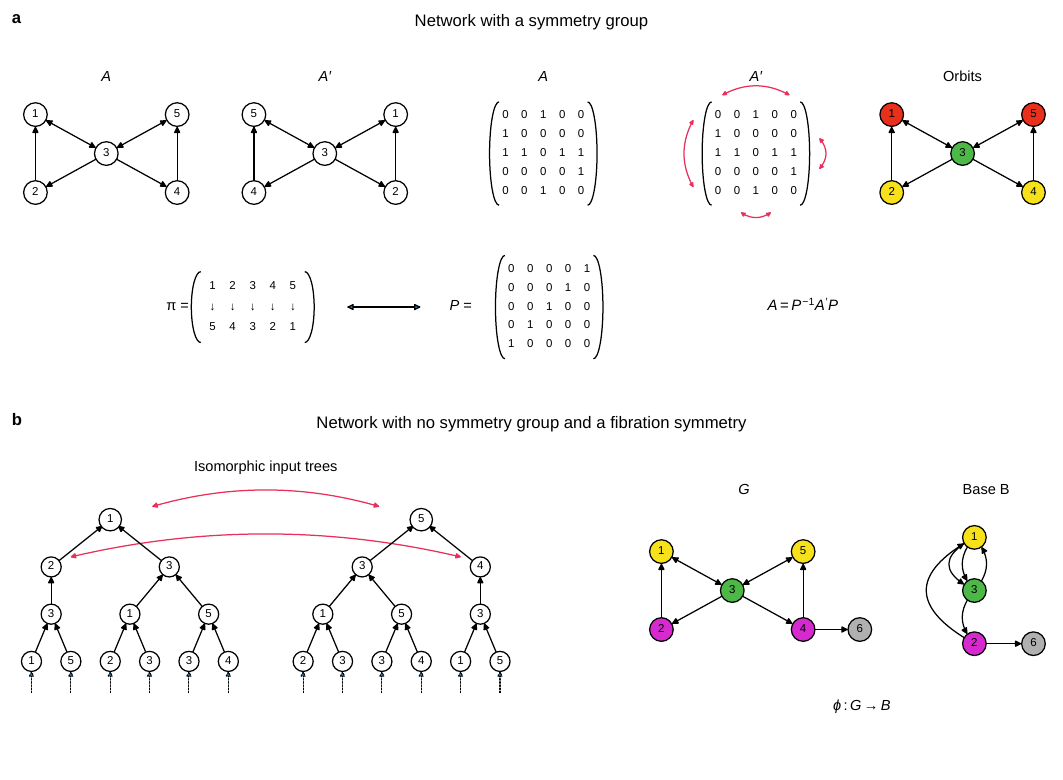}
\caption{{\bf Symmetry formalism.} {\bf a} Example of automorphism in a graph with group symmetry. Left: a permutation $\pi$ transforms the graph with adjacency matrix $A$ into $A'$; this is written in matrix notation through the permutation matrix $P$. The permutation is a symmetry when $A=A'$. Right: applying all symmetries to every node generates the orbital partition shown by the colored nodes. {\bf b} Example of fibration symmetry in a graph with no group symmetry. The addition of the outgoing edge from node 4 to node 6 destroys the global automorphism in {\bf a}. The symmetry fibration remains, since there are nodes (nodes 1 and 5, and nodes 2 and 4) with isomorphic input trees (left). The fiber partition is shown in graph $G$: nodes with the same color (i) belong to the same fiber, (ii) are balanced because they receive the same colors from their in-neighbors, and (iii) are synchronized under any admissible dynamics. The fibration $\phi$ collapses the fibers into the base $B$ following the lifting property (right).}
\label{fig:symmetry}
\end{figure*}

Classically, symmetries are mathematically captured by automorphisms \cite{dixon1996permutation}. In a graph, an automorphism is a permutation of the nodes that preserves the global adjacency connectivity (Fig.~\ref{fig:symmetry}a and Methods Sect.~\ref{sec:automorphism}): the incoming and outgoing neighbors of every node are preserved by the permutation. This is a global condition, because the map involves all nodes and must preserve the entire adjacency matrix. Clusters of nodes mapped into each other by these permutations are called orbits, and nodes within each orbit synchronize their activity under any dynamical system admissible for the network. Because the condition is global, orbits are difficult to obtain in real networks.

A graph homomorphism called graph fibration \cite{boldi2002fibrations,morone2020fibration} defines less constrained, local symmetries (Methods Sect.~\ref{sec:fibration}). Graph fibrations derive from the categorical notion with the same name introduced by Grothendieck and others in the 1960s \cite{grothendieck1959technique}. A fibration is a graph homomorphism that is only required to be a bijection of local in-neighborhoods rather than of the entire network. An equivalent definition, based on the input tree of a node, was given in ref.~\cite{morone2020fibration} (Fig.~\ref{fig:symmetry}b, left). The input tree of a node $v$ contains all the paths that terminate on $v$, and thus represents the history of the information flow reaching $v$ through the network. A fibration $\phi$ of $G$ collapses the nodes of $G$ with isomorphic input trees into the base $B$ (Fig.~\ref{fig:symmetry}b, right). The fibration that produces the minimal base, that is, the maximal collapse, is the symmetry fibration of $G$, since it collects all the symmetries of the network \cite{morone2020fibration}. Sets of nodes with isomorphic input trees are called fibers (the colored nodes in Fig.~\ref{fig:symmetry}b), and play in the fibration framework the role that orbits play for automorphisms. Equivalently, the fiber partition is the minimal balanced coloring of the network, in which nodes of the same color receive the same number of inputs from each color (Methods Sect.~\ref{sec:fibration}). (The term fiber in the fibration sense should not be confused with the white-matter fibers reconstructed by fiber tracking.)

DeVille and Lerman~\cite{deville2015modular} (Theorem 4.3.1 and Lemma 5.1.1; see also refs.~\cite{nijholt2016graph,golubitsky2006nonlinear}) proved that the nodes of a fiber are synchronous under any dynamical system admissible with the graph, independently of the specific dynamics. This result links the fibration of the graph (structure) to the existence of CS (function), and is the basis of the analysis that follows.

\subsection{Functional network of synchrony between ROIs}
\label{sec:synchro_condition}

We analyze BOLD fMRI signals from $n=20$ healthy volunteers with no neurological or psychiatric morbidities in three conditions: at rest \cite{azeez2017review} and while performing a letter fluency and a verb generation task \cite{li2020core} (Methods Sect.~\ref{sec:languagetask}). During the letter fluency task, participants silently generate as many nouns as possible starting with a given letter. During the verb generation task, they generate action words associated with the presented nouns. During the resting state, they lie in the scanner with their eyes open, try not to think of anything in particular, and fixate on a central cross on a screen.

The functional network is built between ROIs primarily involved in language according to the dual stream model \cite{hickok2007cortical, hickok2022dual} (Fig.~\ref{fig:intro}a and Methods Sect.~\ref{sec:dual}). Specifically, we include clinically relevant regions \cite{li2017lexical,li2019functional,li2020core,li2021monolingual} of the dorsal stream: supplementary motor area (SMA), premotor area (PreMA, left and right), supramarginal gyrus (SMG, left and right), Broca's area (BA, left and right), angular gyrus (AG, left and right) and Wernicke's area (WA, left and right). Many secondary areas are also involved in language; we restrict the analysis to this core set (see Discussion).

For each participant, we compute the phase locking value (PLV) \cite{bruna2018phase}, corrected against surrogate data, between the BOLD time series \cite{bullmore2009complex} of each pair of ROIs as the synchronization measure (\SuppFig{plv} and Methods Sects.~\ref{sec:experiments} and \ref{sec:synchronization}), obtaining a functional synchronization matrix for each participant (\SuppFig{correlation}). Averaging these matrices over the $n$ participants gives a weighted group-average matrix with entries in $[0,1]$, from which the CS of ROIs are obtained.

Ideally, a CS is a fully connected, non-overlapping induced subgraph (clique) of the functional network. Since this ideal synchronization cannot be expected from real data, we relax the condition by allowing the clique to be connected to the rest of the network by weaker links. We define a CS $N$-clique as the induced, fully connected subgraph of the functional network composed of $N$ nodes that satisfy
\begin{multline}
\sum_{i<j}^{1,N} \sigma(x_i (t),x_j(t)) \geq {N(N-1)\over
  2}\sigma(x_k (t),x_{k'}(t)) \\ \forall  k= 1,...,
N\mbox{ and } k'  \in \mathcal{M}_k ,
\label{eq:1}
\end{multline}
where $\mathcal{M}_k$ is the set of nearest neighbors of node $k$ that do not belong to the clique or to a clique with higher Cluster PLV (the average PLV within the clique), and $\sigma(x_i(t),x_j(t))$ is the PLV of the time series $x_i(t)$ and $x_j(t)$ of nodes $i$ and $j$ (Methods Sect.~\ref{sec:synchronization}).

Candidate clusters are identified with a percolation procedure \cite{gallos2012small, mastrandrea2023information} on the synchronization matrix (\SuppFig{perolation_cliques}). Starting from a disconnected graph, links are added in decreasing order of weight (synchronization), and the groups of ROIs that become connected along this sequence are taken as candidate cliques. A partition of the ROIs into non-overlapping candidate cliques is accepted as the CS partition when, ordering the cliques by decreasing Cluster PLV, each clique satisfies Eq.~(\ref{eq:1}), so that every clique is compared only with ROIs that are not already assigned to a more strongly synchronized clique. The ordering by Cluster PLV defines a hierarchy of CS, which segregates the global synchronization pattern into locally fully connected synchronized groups.

\subsection{Cluster synchronization at rest and during language tasks}
\label{sec:synchro_rest}

Figure~\ref{fig:functional}a,b shows the resting-state (RS) group-average PLV matrix and the corresponding functional network, with the CS of Eq.~(\ref{eq:1}) shown in different colors (Methods Sect.~\ref{sec:restingtask}). The RS functional network is known to be approximately left-right symmetric \cite{teghipco2016disrupted,seitzman2019state}. Consistently, we find three bilateral CS, each formed by a homotopic pair of regions (SMG, AG and BA). In addition, the largest CS (LCS) is formed by two bilateral pairs (PreMA and WA) together with the SMA. The LCS is consistent with the auditory-motor integration mechanism of the dorsal stream of language processing (Methods Sect.~\ref{sec:restingtask}).

A different functional network emerges during the language tasks. In both letter fluency (Fig.~\ref{fig:functional}c,d) and verb generation (Fig.~\ref{fig:functional}e,f), left lateralization \cite{friederici2011brain} appears through the involvement of the left BA, which joins the SMA. The left BA is recruited by the SMA and desynchronizes from the right BA, which in turn synchronizes with the left and right WA. The set of CS is identical in the two tasks (Fig.~\ref{fig:functional}c,e): the SMA with the left BA, which forms the most synchronized cluster in both tasks, the bilateral PreMA, the bilateral AG, the LCS, formed by the left and right WA and the right BA, and the bilateral SMG. These results are consistent with current models of language processing (Methods Sect.~\ref{sec:restingtask}).

\subsection{Symmetries of the dual stream baseline connectome of language}
\label{sec:highway}

Having identified the CS of the functional network, we look for the corresponding fibers in the connectome of white-matter tracts between the ROIs \cite{dick2014language} (Fig.~\ref{fig:intro}a, right, and Methods Sect.~\ref{sec:structural}). The tracts of the dorsal stream belong to the superior longitudinal fasciculus (SLF) and arcuate fasciculus (AF) system. The AF connects the inferior frontal gyrus (Broca's area) to the posterior superior temporal gyrus (Wernicke's area); the SLF connects Broca's area and the premotor area to the inferior parietal areas (supramarginal and angular gyri) \cite{chang2015contemporary,dick2014language}. The frontal aslant tract (FAT) connects Broca's area with the supplementary motor area and serves the verbal fluency components of language \cite{catani2013novel,jenabi2014}. Sensorimotor integration culminates in Broca's area and the ventral PreMA, which are responsible for articulatory planning \cite{lyo2015pre,rech2019probabilistic}. Two parallel dorsal pathways connect the PreMA (dorsal pathway I) and Broca's area (dorsal pathway II) to Wernicke's area, the first predominantly supporting sound-to-motor mapping and the second higher-level language processes \cite{friederici2011brain}.

These tracts form the dual stream (dorsal) baseline connectome of language (Fig.~\ref{fig:intro}a). A symmetry analysis of this connectome (Fig.~\ref{fig:broken}a and Methods Sect.~\ref{sec:comparison_connectome}) shows that its numbers of orbits and fibers coincide (five), so that the automorphisms (global symmetry) and the fibration symmetries (local symmetry) of this network are the same. The five fibers, which are also orbits, are (\SuppFig{comparison_connectome}): a four-ROI cluster formed by the left and right WA and the left and right SMG, and four clusters that respect the left-right symmetry: PreMA (left and right), BA (left and right), AG (left and right) and SMA (alone). For example, the input trees of WA$_{\mathrm{L}}$, WA$_{\mathrm{R}}$, SMG$_{\mathrm{L}}$ and SMG$_{\mathrm{R}}$ are isomorphic (\SuppFig{comparison_connectome}c), so these ROIs belong to the same fiber. At the same time, the permutation that maps WA$_{\mathrm{L}}$ to SMG$_{\mathrm{L}}$ and WA$_{\mathrm{R}}$ to SMG$_{\mathrm{R}}$ (\SuppFig{comparison_connectome}b), written in cycle notation as $\pi_2 = ({\rm WA}_{\mathrm{L}}\, {\rm SMG}_{\mathrm{L}}) \, ({\rm WA}_{\mathrm{R}}\, {\rm SMG}_{\mathrm{R}})$, is an automorphism, so the same four ROIs also form an orbit (red in Fig.~\ref{fig:broken}a). This coincidence of fibers and orbits is a condition of high symmetry of the set of inter-regional routes available for language processing.

A synchronization pattern that emerges during functional engagement (Fig.~\ref{fig:functional}) need not be supported by the full symmetry of the baseline connectome. Lower symmetry corresponds to two distinct situations: the number of orbits exceeds the number of fibers, indicating a loss of global group symmetry, or the number of fibers increases, indicating a loss of local fibration symmetry. We examine both cases below.

\subsection{Inferring the structural network sustaining rest and language from cluster synchronization}

\begin{figure*}[t!]
\centering \includegraphics[width=\textwidth]{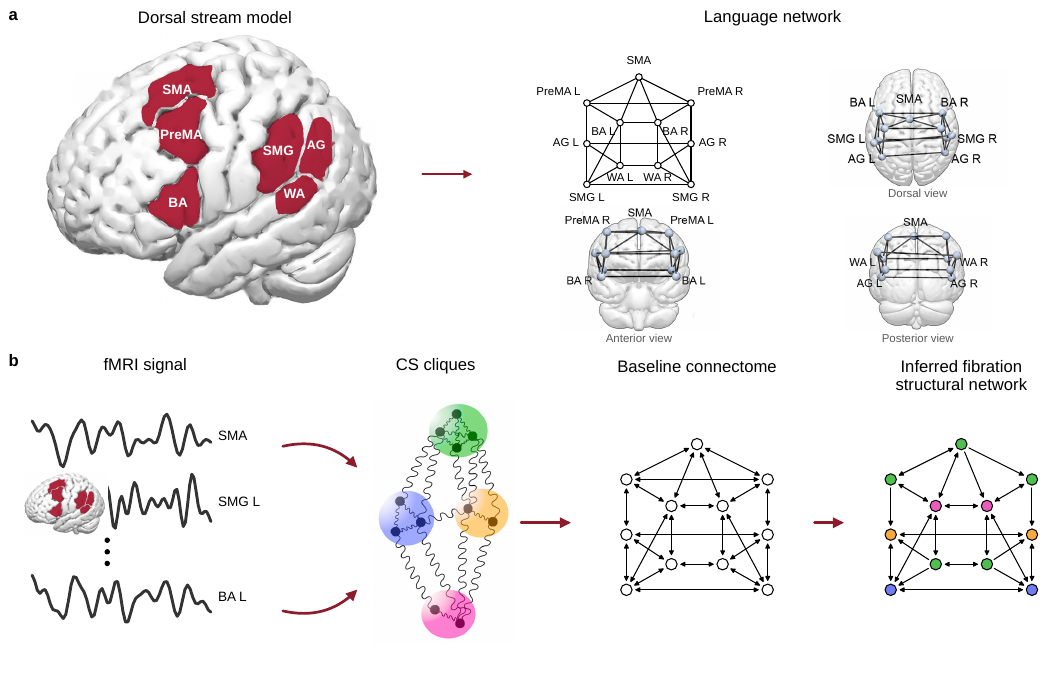}
\caption{{\bf Dual stream model of language and the inference scheme.} {\bf a} Left: regions of interest (ROIs) of the primary language network given by the dorsal stream of the dual-stream model, shown on a three-dimensional rendering of the brain. Right: dual (dorsal) stream baseline connectome showing the white-matter tracts between the ROIs. {\bf b} Pipeline for the inference of the structural network from cluster synchronization (CS) data. Left: functional magnetic resonance imaging (fMRI) data acquired in the resting state (RS) or during a language task in $n=20$ participants (shown: z-scored RS blood-oxygen-level-dependent time series of three ROIs in one participant) are used to compute the group-average CS cliques among ROIs. The CS are identified with the colors in the baseline connectome. A mixed integer linear programming algorithm is used to infer the structural network (right) that sustains the cluster pattern obtained from the dynamics. ROI abbreviations: SMA, supplementary motor area; PreMA, premotor area; SMG, supramarginal gyrus; BA, Broca's area; AG, angular gyrus; WA, Wernicke's area; L and R, left and right hemisphere. Source data are provided as a Source Data file.}
\label{fig:intro}
\end{figure*}

The baseline connectome represents the set of available routes between the ROIs, but which of these routes are used depends on the state of the brain \cite{park2013structural}. According to the hypothesis formulated in ref.~\cite{park2013structural}, the brain uses a subset of the links of the connectome in each functional state. This one-to-many structure-function relation allows diverse functional states (rest, language, etc.) to emerge from a single static connectome. In the present case, different subsets of the dorsal stream connectome of Fig.~\ref{fig:intro}a should mediate different functional networks \cite{park2013structural,friston2011functional}.

To test this hypothesis, we infer the state-specific structural subgraph associated with the balanced coloring of each functional network observed experimentally. We develop a mixed integer linear programming (MILP) algorithm \cite{bertsimas1997introduction,leifer2022symmetry} that removes the minimal number of links from the connectome so that its fibers match the coloring obtained from the data. The hypothesis is falsifiable: if it holds, the MILP must find a solution using only link removals; if no such solution exists, the hypothesis is rejected. The inference proceeds in three steps (Fig.~\ref{fig:intro} and Methods Sect.~\ref{sec:algorithm}):
\begin{itemize}
    \item for the chosen set of ROIs (Fig.~\ref{fig:intro}a, left), define the baseline connectome, that is, the graph of all allowed structural connections among them (Fig.~\ref{fig:intro}a, right);
    \item find the CS of the functional network for a given state with Eq.~(\ref{eq:1}) (Fig.~\ref{fig:functional}a,c,e), and assign to the ROIs of each CS a color, which defines the target balanced coloring;
    \item remove the minimal number of edges from the baseline connectome until the fibers of the resulting graph match the target coloring (Fig.~\ref{fig:intro}b, right).
\end{itemize}

We apply the algorithm to the functional networks at rest and during the two language tasks. In all cases, the MILP finds a solution using only link removals, consistent with the one-to-many hypothesis (10 of the 46 directed links of the connectome are removed for the resting state and 14 for the tasks). Since the coloring is the same in the two tasks, the same structural subgraph supports both tasks.

\subsection{Symmetry-breaking transition to resting state and task}

Much of the complexity of the natural world emerges from mechanisms of symmetry breaking \cite{anderson1972more,yeomans1992statistical} (Methods Sect.~\ref{sec:breaking}). A physical system exhibits spontaneous symmetry breaking when its ground state, the state of minimum energy, has less symmetry than the system itself. Different phases of matter are characterized by different symmetries: at higher temperatures, matter takes on a higher-symmetry phase (for example paramagnetism, normal conductivity or fluidity), whereas at lower temperatures the symmetry is broken (ferromagnetism, superconductivity or superfluidity).

The connectome is not a dynamical state, but the transition from the baseline connectome to the subgraphs that sustain rest and task can be described in analogy with symmetry breaking. The symmetry of each network is quantified by the number of orbits (global symmetry) and the number of fibers (local symmetry) of its minimal coloring: fewer classes correspond to higher symmetry, in the same way as a non-zero magnetization signals reduced symmetry in a ferromagnet. With these two measures the transitions form a progression of symmetry breaking in two steps. From the baseline connectome to the resting state, the global symmetry is broken (the number of orbits increases from five to eleven) while the left-right fibration symmetry is retained (all four resting-state fibers are bilateral, compared with five fibers in the baseline connectome). From the resting state to the tasks, the local symmetry is broken as well (the number of fibers increases from four to five), with the global symmetry remaining broken. The control parameter is the functional state (rest or task), which acts through the structural modulation inferred by the MILP; with the present experimental design it is discrete, and parametrically graded task demands could support a continuous treatment in future work. A change in the number of fibers is a necessary but not sufficient signature of symmetry breaking in the physical sense, which also requires the reduced-symmetry configuration to be a dynamically stable state. We test this in Methods Sect.~\ref{sec:kuramoto}: Kuramoto oscillator dynamics, admissible in the sense of refs.~\cite{golubitsky2006nonlinear,stewart2003symmetry,pecora2014cluster,sorrentino2016complete}, simulated on the inferred resting-state subgraph reproduce the resting-state CS pattern in almost all realizations in the weak-coupling regime. The fiber partition therefore corresponds to a synchronization state that the dynamics reaches robustly; the five-node fiber, whose members are not all directly connected, reaches it after the longest transients.

Starting from the baseline connectome (Fig.~\ref{fig:broken}a), we find two successive symmetry-breaking steps as the brain engages in different states (Fig.~\ref{fig:broken}b,c). The first occurs at rest. Figure~\ref{fig:broken}b shows the balanced coloring of the structural subgraph inferred for the resting state. Its symmetry analysis (Methods Sect.~\ref{sec:comparison_rs} and \SuppFig{comparison_rs}) shows that the group symmetry, including the global left-right symmetry, is lost: the subgraph has only the trivial automorphism, with one orbit per ROI (\SuppFig{comparison_rs}a, right, and b). Fibration symmetry instead increases, with four fibers at rest (four colors in Fig.~\ref{fig:broken}b) compared with five in the baseline connectome (\SuppFig{comparison_rs}a, left, and c). Thus, although the global left-right symmetry is broken, the local left-right fibration symmetry required for bilateral synchronization is retained. The resting-state synchronization is therefore compatible with a structural subgraph that loses the automorphisms of the baseline connectome while retaining its left-right fibration symmetry. The fiber partition is reorganized rather than simply coarsened: WA and SMG, which share a fiber in the baseline connectome, belong to different fibers at rest.

The second step occurs during the language tasks. Figure~\ref{fig:broken}c shows the balanced coloring of the structural subgraph inferred for the tasks (Methods Sect.~\ref{sec:comparison_task} and \SuppFig{comparison_task}). The lateralization of activity breaks the fibration symmetry between the left and right BA, which now belong to two different fibers: the left BA is recruited by the SMA, whereas the right BA is recruited by the left and right WA, which remain locally symmetric. The number of fibers increases from four to five, indicating lower fibration symmetry than at rest, as if the task acted as a perturbation of the resting-state dynamics. The global symmetry remains broken, with only the trivial automorphism and one orbit per ROI (\SuppFig{comparison_task}a, right, and b).

The five task fibers and their lateralization are compatible with neurocognitive models of language production. Networks involving the temporal and inferior frontal cortices, predominantly lateralized to the left hemisphere, support syntactic processes, whereas less lateralized temporo-frontal networks support semantic processes \cite{friederici2013language, roger2022unraveling}. Symmetry breaking is therefore consistent with the specialization of the SMA, BA and WA for specific, mainly syntactic, operations, as observed in other brain regions that display functional asymmetry \cite{toga2003mapping}. Overall, these results show that different synchronization patterns can be matched to different structural subgraphs identified by symmetry considerations, so that the mesoscopic correspondence between structural and functional connectivity can be described as a reconfiguration process constrained by fibration symmetry.

\section{Discussion}
\label{sec:discussion}

A central challenge in cognitive neuroscience is the structure-function mapping problem: how static anatomical connections support dynamic functional interactions. Prior studies have demonstrated statistical relationships between structural and functional connectivity \cite{honey2009, greicius2009resting, hermundstad2013structural, goni2014resting}, but how cognitive demands shape functional networks in a principled way is still an open question. Here we show that cluster synchronization among regions involved in language can be accounted for by local symmetry properties of the structural connectome \cite{bullmore2012economy,park2013structural}. Using resting-state and task fMRI data, we find that in each state the synchronized language regions can be matched to the fibers, defined by the isomorphism of the input trees, of a subgraph obtained from the baseline connectome by link removals only. This match is not guaranteed a priori: had no deletion-only solution existed, the hypothesis would have been rejected. According to fibration symmetry theory \cite{deville2015modular}, these groups are dynamical modules that synchronize because they share structural input profiles. This result helps clarify how the brain supports lateralized and distributed language functions, extending models such as the dual-stream framework \cite{hickok2007cortical} with network-level dynamics.

The changes in network configuration that accompany the transition from rest to task can be described as symmetry-breaking transitions, similar to phase transitions in physical systems \cite{anderson1972more,yeomans1992statistical,goldenfeld2018lectures}. At rest, the symmetry of the structural subgraph supports bilateral synchronization, consistent with the high inter-hemispheric coupling observed in resting-state networks \cite{biswal1995functional, damoiseaux2006consistent}. During language tasks such as verb generation and letter fluency, this symmetry is broken by the recruitment of the left Broca's area and related motor regions \cite{price2012review, fedorenko2014reworking}, consistent with the established lateralization of language in functional connectivity \cite{knecht2000handedness, toga2003mapping}. Our results extend this picture from the functional to the structural network: lateralization corresponds to a breaking of fibration symmetry in the subgraph of structural connections that sustains the task.

The MILP approach makes this transition visible in the connectome by identifying the subnetworks specific to rest and task. It operationalizes the communication-subset hypothesis, according to which only a portion of structural connections is functionally relevant at a given time \cite{deco2011emerging,sporns2013network}, and provides a concrete method to infer which connections these are. The methodology can be applied to any cognitive (multimodal) or non-cognitive (unimodal) task: each task induces a specific pattern of cluster synchronization, sustained by a structural subgraph that can be inferred through fibration symmetries. We therefore expect other, task-specific patterns of synchrony to emerge in non-language tasks.

The symmetry-breaking framework also connects with the developmental trajectory of language lateralization across the lifespan \cite{olulade2020neural}. Language processing is relatively bilateral in early childhood and becomes increasingly left-lateralized with maturation and cognitive specialization, a progressive symmetry-breaking process. This trajectory is not unidirectional: in old age, language networks often return partially to a more bilateral engagement, consistent with compensatory or reorganizational processes. From this perspective, fibration symmetry breaking offers a common framework to describe how brain networks move between flexible, redundant states and more specialized configurations during development and aging. Longitudinal studies could test whether the number of fibers in the language network changes systematically across developmental stages, providing a quantitative measure of this trajectory.

By connecting formal symmetry principles with empirical neuroimaging data, the framework may be useful for studying language reorganization after injury \cite{duffau2008anatomo}, identifying neuromodulation targets \cite{bassett2017network}, and improving models of language acquisition and processing in health and disease. The study of the controllability of brain networks \cite{gu2015controllability}, an open problem in neuroscience, may also benefit from these results. Invasive (surgical) and non-invasive (electric or magnetic stimulation) treatments of neurological and psychiatric diseases \cite{li2020core} could use the identification of symmetry and synchronization patterns, and of their breaking, to reduce side effects or improve effectiveness.

The present study characterizes the symmetry-breaking pattern associated with functional state transitions, but does not identify the neural mechanisms that drive these transitions. Which mechanisms, whether top-down control from prefrontal regions, neuromodulatory signals or task-driven external input, selectively modulate the connections that break fibration symmetry remains an open question. The fibration framework, however, turns this question into a constrained one: rather than asking generically what drives task-related reconfiguration, one can ask which connections must be modulated to move between two fiber partitions, and which neural processes achieve this modulation. This formulation makes the symmetry-breaking hypothesis directly testable, for example through targeted neuromodulation or computational modeling of input-driven dynamics on the inferred structural subgraphs.

The 11-ROI language network analyzed here was chosen as a focused and interpretable proof of concept. Both tasks involve processes beyond the dorsal stream: verb generation involves auditory comprehension and semantic search, which rely on the ventral stream, and letter fluency involves orthographic-phonological mapping. We focused on the dorsal stream to capture the primary articulatory, motor-speech and auditory-motor integration pathways. The small network size also allows an exact algebraic symmetry analysis and a direct comparison with established neurocognitive models. The fibration framework itself is not limited to small networks: it has been applied to the whole {\it C. elegans} connectome (302 neurons) \cite{avila2024fibration}, to genetic regulatory networks \cite{stewart2024dynamics} and to metabolic networks \cite{alvarez2025symmetries}, and algorithms for computing balanced colorings scale polynomially with network size. The inference of the structural subgraph by integer programming is NP-hard in the worst case and has so far been applied to circuits of a few tens of nodes \cite{avila2025symmetries}; extending it to larger networks will require heuristic or decomposition strategies. A weighted extension of the analysis, based on an empirical diffusion-tractography connectome and accounting for known structural asymmetries, is reported in Methods Sect.~\ref{sec:weighted}; incorporating the ventral stream remains a direction for future work.

\begin{figure*}[t!]
\centering
\includegraphics[width=\textwidth]{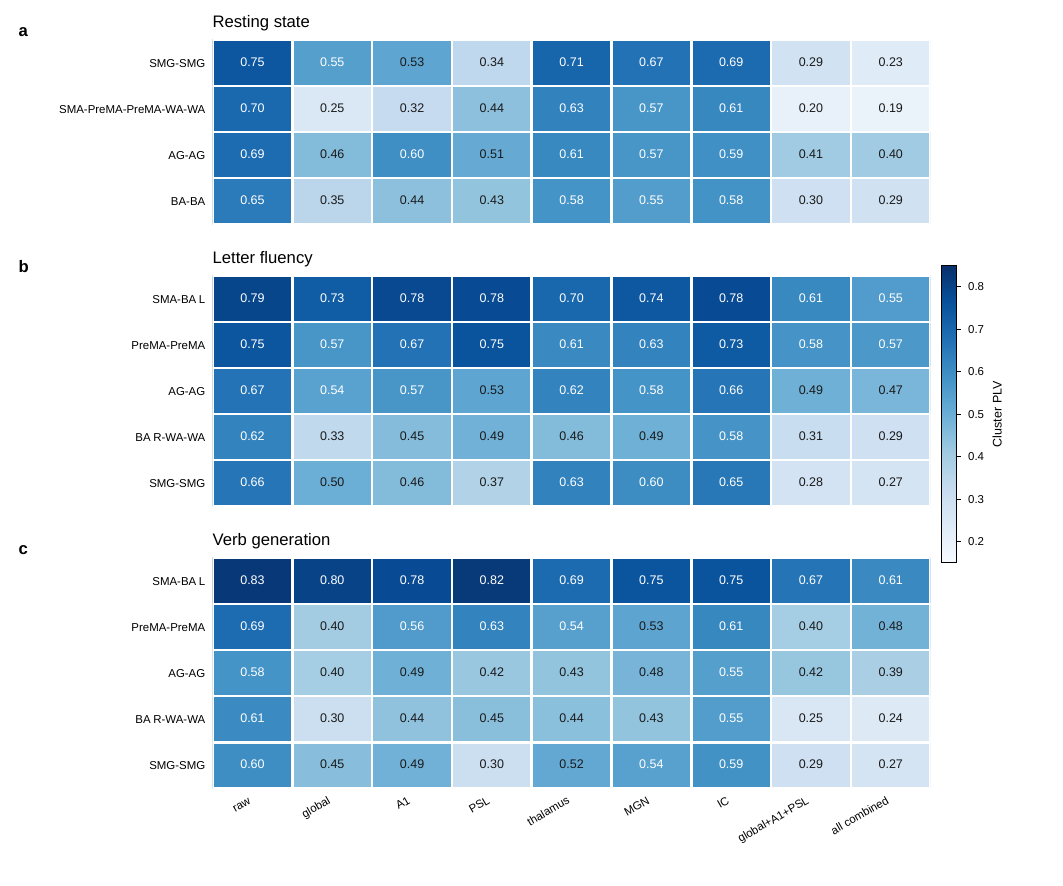}
\caption{{\bf Whole-brain control for shared or external input in the resting state and in two task-fMRI datasets.} Cluster phase locking value (Cluster PLV, the surrogate-corrected PLV averaged within each clique; numbers) for the cliques of the CS partitions reported for the resting state and the tasks, across the nine whole-brain control conditions (Methods Sects.~\ref{sec:wholebrain} and \ref{sec:task-generalization}; Tables~\ref{tab:wholebrain-rs}--\ref{tab:suppl-vg}), in the resting state ({\bf a}), letter fluency ({\bf b}) and verb generation ({\bf c}) tasks ($n=20$ participants in each dataset). All cliques satisfy Eq.~(\ref{eq:1}) in every condition, within the PLV tolerance of 0.01 of the control analyses. Conditions: raw, unregressed data; global, mean cortical signal; A1, primary auditory cortex; PSL, perisylvian language area; MGN, medial geniculate nucleus; IC, inferior colliculus; combined, simultaneous regression of the indicated signals. X-X denotes the left-right homotopic pair of region X. ROI abbreviations: SMA, supplementary motor area; PreMA, premotor area; SMG, supramarginal gyrus; BA, Broca's area; AG, angular gyrus; WA, Wernicke's area; L and R, left and right hemisphere. Source data are provided as a Source Data file.}
\label{fig:wholebrain_heatmap}
\end{figure*}

Finally, the brain is highly interconnected, and connections outside the ROI set may contribute to the synchronization patterns observed within the language network. The existence of MILP solutions that use only edge removals supports the sufficiency of the within-network architecture, but does not exclude external contributions. We tested this possibility with whole-brain fMRI data from all 20 participants (Methods Sect.~\ref{sec:wholebrain}). Regressing candidate shared signals (global cortical signal, primary auditory cortex, perisylvian language area, thalamus, medial geniculate nucleus and inferior colliculus) out of the 11 ROI time series before recomputing the CS reduces the Cluster PLVs but leaves all four resting-state clusters valid under Eq.~(\ref{eq:1}) in every condition. The largest reductions follow the removal of cortical signals, and the SMG cluster is selectively reduced by the signal of the adjacent perisylvian language area, consistent with the spatial proximity of the two regions, whereas the subcortical regressors change the Cluster PLVs only modestly, indicating that the shared component of the pattern is cortical rather than subcortical. Applying the same control to the two task datasets (Methods Sect.~\ref{sec:task-generalization}) shows that all five task clusters remain valid in every condition in both tasks (within the PLV tolerance of 0.01 adopted for the control analyses), and that during verb generation the dominant fronto-opercular cluster (SMA and left BA) remains the most synchronized clique in every condition, including the combined-signal control (Fig.~\ref{fig:wholebrain_heatmap}).

Overall, our findings suggest that local symmetry at the mesoscopic level constrains how brain structure supports coherent function. Symmetry fibrations generalize the symmetry groups of physics and have been found in biological systems ranging from the {\it C. elegans} connectome to genetic and metabolic networks. Symmetry fibrations may therefore provide a common descriptive framework for physical and biological networks, extending the use of symmetry principles to biological systems.

\newpage

\section*{Methods}

This research complies with all relevant ethical regulations. The study protocol was approved by the Institutional Review Board of Memorial Sloan Kettering Cancer Center (MSKCC), New York, USA (protocol IRB-18-121), and the study was conducted in accordance with the Declaration of Helsinki. Written informed consent was obtained from all participants.

\section{Experimental protocols}
\label{sec:experiments}

\subsection{Subjects.}

Twenty healthy right-handed adults (13 men and 7 women; mean age 36.6 years, standard deviation 11.6 years; age range spanning 36 years) without any neurological or psychiatric history participated in the study. Sex was self-reported by the participants. Sex and gender were not considered in the study design, and no sex- or gender-based analysis was performed because the sample does not provide adequate statistical power for subgroup comparisons; sex-disaggregated participant numbers are reported here. Participants were volunteers and were compensated by MSKCC for their participation. No statistical method was used to predetermine the sample size. The sample size is comparable to that of previous functional magnetic resonance imaging (fMRI) studies of language mapping with the same tasks \cite{li2020core,li2021monolingual}, and the inter-subject variability of the group-averaged synchronization matrices is quantified in Methods Sect.~\ref{sec:synchronization}.

\subsection{MRI methods.}

A GE 3T scanner (General Electric, Milwaukee, Wisconsin, USA) and a
standard quadrature head coil were used to acquire MR
images. Functional images covering the whole brain were acquired using
a (T2*)-weighted imaging sequence sensitive to the blood oxygen
level-dependent (BOLD) signal (repetition time, TR/TE =
2500/40 ms; slice thickness = 4.5 mm; matrix = 128 $\times$ 128; FOV =
240 mm; volumes = 160). Functional matching axial T1-weighted images
(TR/TE = 600/8 ms; slice thickness = 1 mm) were acquired for
anatomical co-registration purposes.

\subsection{Language tasks and RS.}
\label{sec:languagetask}

Three different fMRI acquisitions, each lasting 400 seconds (160 volumes $\times$ TR = 2.5 seconds), were performed under different experimental conditions:  resting-state, verbal fluency task
using verb generation in response to auditory nouns, and letter fluency letter task in response to visual task instructions.

During the resting state condition, subjects are asked to lie in the
scanner and keep their eyes open, to try to think of nothing in
particular, and to keep fixating on a central cross on a screen during RS.

In the verb generation task, subjects were
presented with a noun by oral instruction and then asked to generate
verbs associated with the noun.  For example, subjects are
presented with a noun (for example, baby) and asked to generate verbs associated with it (for example, cry or crawl). Subjects
perform the task silently to avoid motion artifacts.  Four nouns are
presented aurally in each of eight stimulation epochs, each lasting 50 s,
for a total of 32 distinct nouns over the entire duration. Each
epoch consisted of a resting period (30 s) and a task period (20 s).

In the letter fluency task, on the other hand, subjects are asked to generate nouns
that begin with a given letter silently. For example, the subject 
presented with the letter A can generate words such as apple, apron or ashtray.  Stimuli are displayed on a screen over eight
stimulation epochs, each lasting 20 s. During the task, two
letters are presented in each stimulation epoch. Each epoch also
consisted of a 30-second resting period during which subjects were
asked to focus on a blinking crosshair.

In order to avoid artifacts from jaw movements, subjects were asked to silently generate the words. Although subtle differences in movement between rest and task could potentially explain differences in measures of signal coherence, we note that the images were acquired at MSKCC by trained personnel following standardized clinical acquisition protocols.  According to the complete motion control and preprocessing pipeline described in Section \ref{sec:datapreproc}, possible differences in signal coherence are unlikely to be due to
motion artifacts.
Finally, we note that response exhaustion, a known phenomenon in clinical fluency tasks, could in principle affect functional dynamics measured during task blocks. However, the relatively short block durations used here (20 s) mitigate this concern. Moreover, our analysis is based on time-averaged phase synchronization (PLV), which integrates over within-block temporal dynamics and is therefore robust to moderate fluctuations in response rate.

\subsection{Data preprocessing.}
\label{sec:datapreproc}

Functional MRI data were processed and analyzed using the software
program Analysis of Functional NeuroImages (AFNI, version 18.2.15) \cite{cox1996afni}. Head
motion correction was performed using 3D rigid-body registration. The
first volume was selected to register all other volumes. The first
volume was chosen because it was acquired prior to anatomical
scan. Task and resting state fMRI scans were monitored using real time post-processing software BrainWave (BrainWave RT, Medical Numerics, Germantown, MD) to monitor brain activity and head motion. For subjects showing severe head motion over time, the scan is generally repeated. For small head motion (less than 2 voxel size), a motion correction algorithm (iterated linearized weighted least squares) was applied considering 3 translation and 3 rotation parameters against a reference volume.  The obtained 6 parameter time courses were also integrated in the statistical analysis to regress out residual motion-correlated artifactual voxels.
Spatial smoothing was applied to improve the
signal-to-noise ratio using a Gaussian filter with a 6 mm full width of
half maximum. Corrections for linear trend and high-frequency noise
were also applied. The Resting-state data requested some more
preprocessing steps. The first five volumes were discarded to allow the longitudinal magnetization to reach a steady state (T1 equilibration), leaving 155 volumes for the analysis. The data were corrected for head motion by
regressing head motion data and the first five principal
components of the white matter and CSF signals. They were also
detrended, demeaned, and band-pass filtered (frequency range 0.01-0.1
Hz).  All fMRI data were registered in the standard space (Montreal
Neurological Institute MNI152 standard map). 

\section{Definition of ROIs: dorsal stream model of language}
\label{sec:dual}

Language processing modeling has long been based on 
the Geschwind-Lichtheim-Wernicke model \cite{geschwind1967wernicke},
 drawn primarily from observations of individuals with brain injuries.
Following this model, words are perceived through a dedicated word
reception center (Wernicke's area) within the left temporoparietal
junction. Subsequently, this region sends signals to a word production
center (Broca's area) in the left inferior frontal gyrus.

Advancements in electrophysiological and magnetic resonance imaging techniques have revealed
a dual auditory pathway. This led to the development of a dual stream
model \cite{hickok2007cortical, hickok2022dual}. According to this
model, two distinct pathways connect the auditory cortex
to the frontal lobe, each serving different linguistic functions. The
auditory ventral stream pathway is responsible for sound recognition
and is referred to as the auditory what pathway. On the other hand,
the auditory dorsal stream, found in humans and non-human primates, is
responsible for sound localization and is referred to as the auditory where pathway. In humans, particularly in the left hemisphere, this pathway
also handles speech production, repetition, lip-reading, phonological
working memory, and long-term memory.

The relevant ROIs are those areas involved in the two
language tasks considered \cite{li2020core,li2021monolingual}. Since the tasks are both focused on
language production (letter fluency and verb generation), 
regions of the dorsal stream are part of the analysis
(Fig.\ref{fig:intro}): Supplementary Motor Area (SMA), Premotor
Area (PreMA, left and right), Supramarginal Gyrus (SMG left and
right), Broca's Area (BA, left and right), Angular Gyrus (AG, left and
right), Wernicke's Area (WA, left and right). The BA and WA are
recognized as responsible for the expression and
comprehension of language. The supplementary motor area (SMA) has been widely
considered involved in controlling speech-motor functions, and it has
also been shown \cite{hertrich2016role} that the SMA performs several
higher-level control tasks during speech communication and language
comprehension. Even if SMA can be divided into two symmetric portions in the left and right hemispheres, 
we consider it to be a single region because of its pivotal role as a single unit orchestrating the functional language network \cite{hertrich2016role,liu2020identifying}. 
The AG is assumed to be a region of the brain
associated with complex language functions (i.e., reading, writing, and interpreting what is written). In contrast, the SMG is involved
in the phonological processing of high-cognitive tasks. Finally,
processing an action verb depends in part on activity in a motor
region that contributes to planning and executing the action named by
the verb. The premotor cortex is known to be functionally involved
in understanding action language \cite{chang2023representing}.

\section{Definition of the dual (dorsal) stream baseline connectome of language}
\label{sec:structural}
 
According to the dual stream model introduced in Methods
Sec. \ref{sec:dual}, it is known that human language is based on two
primary pathways of white-matter: 
the dorsal stream, which is related to sensorimotor integration, and
the ventral stream, which is related to speech comprehension
\cite{dick2014language}. The tracts we are interested in are the
primary tracts of the dorsal stream, which is formed by white matter
tracks of the superior longitudinal fasciculus (SLF) arcuate
fasciculus (AF) system. The AF connects the inferior frontal gyrus
(BA) to the posterior superior temporal gyrus (WA); the SLF connects
BA and PreMA to the inferior parietal areas (SMG and AG)
\cite{chang2015contemporary,dick2014language}. The frontal aslant
tract (FAT) connects BA to the SMA, serving the verbal fluency
components of the language \cite{catani2013novel}.  Sensorimotor
integration culminates in the BA and ventral PreMA, which are
responsible for articulatory planning
\cite{rech2019probabilistic}. Two parallel dorsal pathways have also
been described \cite{friederici2011brain}. One connects the PreMA
(dorsal pathway I) and BA (dorsal pathway II) to the WA, with the
first predominantly supporting sound-to-motor mapping and the second
supporting higher-level language processes. It is also known that
PreMA represents a crucial speech production hub thanks to its
coupling with the SLF. Preservation of this cortical-subcortical
connection is crucial for speech integrity and represents an
anatomical constraint to cortical plasticity
\cite{van2014limited}. Additionally, homologous right- and left-sided cortical areas are likely connected by the corpus callosum, the main inter-hemispheric commissure, which enables communication between the two cerebral hemispheres \cite{naidich2012imaging}.
The dorsal stream tracts constitute the set of routes that make up the information transfer highway within
the sensorimotor integration network that supports language processing. Figure\ref{fig:intro}a shows the structural connectivity pattern identified by the neuroradiologists among the authors, according to the information available in the literature on the tracts described above. It shows a highly symmetric structure since the automorphisms are the same as the symmetry fibrations seen in Fig.\ref{fig:broken}a.

We note that the baseline connectome is represented as a bidirected graph. The fibration analysis operates on this graph using directed input trees. The graph is binary (unweighted), reflecting the presence or absence of established tracts rather than their quantitative strength. This choice enables a clean algebraic analysis of symmetry and symmetry breaking. We also carried out a weighted extension (Methods Sect.~\ref{sec:weighted}), generalizing the balanced-coloring constraint to weighted in-strength with an explicit numerical tolerance $\varepsilon$ representing measurement uncertainty. A literature-grounded synthetic single-tract asymmetry (the documented leftward volumetric asymmetry of the arcuate fasciculus) shows that none of the three partitions' exactly-balanced ($\lambda=1$) solutions is selectively immune to this specific perturbation: the imbalance it introduces scales as the fractional asymmetry itself for baseline, resting-state, and task alike (Methods Sect.~\ref{sec:weighted}). Under real diffusion-tractography weights from an independent HCP dataset registered to the same atlas used here \cite{rosen2021whole}, all three partitions require only a small tolerance (1.7\% baseline, 1.5\% task, 2.1\% resting state, as a percentage of the strongest single tract weight) to become feasible, well within reported tractography measurement uncertainty \cite{jones2010precision}. We interpret this as evidence for approximate, rather than exact, weighted fibration symmetry (a quasi-fibration in the sense of \cite{boldi2022quasifibrations}), and note that the resting-state partition, being the most symmetric of the three, is correspondingly the most sensitive to residual weight heterogeneity.
Although the major tracts of the dorsal stream exist bilaterally, it is well established that some tracts (e.g., the arcuate fasciculus) show a volumetric leftward asymmetry in the general population. The symmetric baseline adopted here represents an idealized scaffold of the principal routes, enabling the formal analysis of symmetry breaking. We note that not every published, HCP-derived group connectome is suitable for testing hypotheses about these specific tracts: a heavily-thresholded, group-consensus connectome we examined for comparison (Methods Sect.~\ref{sec:weighted}) failed to recover even the strong, well-established left--right homotopic callosal connection of Broca's area, underscoring the rationale for the literature-curated approach adopted here. The symmetry of the baseline is not an assumption required by the method: the fibration framework applies to any graph, symmetric or not, and the symmetry of the baseline reflects the known bilateral existence of the major tracts. Starting from a symmetric baseline makes symmetry-breaking transitions especially transparent and biologically interpretable. 
When a network like this is equipped with state variables and dynamical equations, it becomes a network system of ordinary differential equations (ODEs). Specific features of the dynamics, such as the quantitative value of interaction constants or the frequency of an oscillation, depend on the precise details of the model equations. Here, we focus on more general features that can occur for broad classes of models and systems. Cluster synchronization is an example of such a feature
\cite{golubitsky2006nonlinear,stewart2003symmetry,pecora2014cluster,sorrentino2016complete}.

\section{Cluster synchronization and the functional network}
\label{sec:synchronization}

It is well established that resting-state functional connectivity shares notable similarities with diffusion-derived structural connectivity patterns \cite{honey2009,greicius2009resting}. Preferential coupling between regions at rest may therefore more directly reflect the underlying wiring patterns than task-evoked dynamics, which additionally reflect how structure is dynamically recruited for specific cognitive demands. This motivates the use of RS functional clusters as a principled basis for inferring the minimal structural constraints that support communication, and supports the symmetry hierarchy we observe: the baseline connectome’s high symmetry is partially preserved in the resting state (reflecting its closeness to the anatomical scaffold) and further broken by the task.

BOLD time series were extracted from all voxels in a 6 mm radius sphere centered on the MNI152 target coordinates addressing a specific language parcel of the Glasser atlas \cite{glasser2016multi}. We identified the parcels [44, 6a, SCEF, STSdp, PF, PFm] as ROIs [BA, PreMA, SMA, WA, SMG, AG], respectively \cite{rolls2022human}. We decided to include spheres of the same size to avoid a biased estimation of the signal averaging within ROIs, the Glasser parcels of interest being composed of different numbers of voxels. Each ROI included 123 voxels. To test the robustness of our choice, we compared Resting State results obtained using the spheres with those coming from the same analysis done on resting state time series averaged in the Glasser parcels of the Dorsal Stream of the Language System. With time series averaged within the Glasser parcels, the same four clusters satisfy Eq.~(\ref{eq:1}) (\SuppFig{extended_figure_glasser}).

BOLD signal synchronization has been widely used to describe phase coupling of large-scale brain activity, controlling for amplitude effects \cite{pang2019power, gravel2018phase, bolt2018inter, kringelbach2020brain, honari2021evaluating, honari2022mode, xu2020inter}. Synchronization between pairs of nodes in the language network was estimated as the Phase Locking Value
(PLV) \cite{lachaux1999measuring,ponce2015resting,zhou2020prediction,honari2021evaluating} between the BOLD time series from
pairs of ROIs. Once time series were obtained for the eleven ROIs
included in the study (by spatial averaging the BOLD signal within each ROI at each time point), synchronization was calculated as follows.
Given the BOLD signals $n_u(t)$ and $n_v(t)$ coming from regions $u,v
= 1, ..., N$ ($N = 11$), their instantaneous phases $\phi_{n_u}(t)$
and $\phi_{n_v}(t)$ can be obtained by means of their Hilbert
transform (see \SuppFig{plv}).  The PLV
$r_{uv}$ is then given by:

\begin{equation}
    r_{uv}=|{\langle e^{-j(\phi_{n_u}(t)-\phi_{n_v}(t))} \rangle}_t| ,
\end{equation}
where $j$ is the imaginary unit.

To correct the PLV for the synchronization expected by chance, for each pair of ROIs we generated 5,000 surrogates by randomly shuffling in time the samples of the second time series, recomputed its phase and the PLV with the first time series, and took the average surrogate PLV $\gamma_{uv}$. The synchronization between ROIs $u$ and $v$ is then
\begin{equation}
\sigma_{uv}=\begin{cases} \dfrac{r_{uv}-\gamma_{uv}}{1-\gamma_{uv}}, & r_{uv}\geq\gamma_{uv},\\[4pt] 0, & r_{uv}<\gamma_{uv}, \end{cases}
\label{eq:plv_corrected}
\end{equation}
which is the quantity referred to as PLV in the rest of the paper.

The values $\sigma_{uv}$ were then entered into a $N\times N$ synchronization or PLV matrix. Finally, the PLV matrices were averaged between subjects in
each experimental condition (resting state, letter fluency task,
verb generation task). The functional network is then obtained by
applying a percolation procedure to the group-averaged synchronization matrix, obtaining the CS as
explained in the main text. To control for the inter-subject variability of the PLVs in the three experimental conditions two estimators were used: the PLV Coefficient of Variation (measured as the ratio between the standard deviation and the average across subjects \cite{anthony2003cambridge}) and the Jensen-Shannon divergence \cite{lin1991divergence} of PLVs distributions among subjects. Both estimators are reported in \SuppFig{variationcoefficient} and \SuppFig{jsd}. The CS analysis is performed on the group-averaged matrices, whose standard error across participants is about 0.05.  

The PLV of the cluster shown in Figures \ref{fig:functional}a, \ref{fig:functional}c, and \ref{fig:functional}e is the
value of the weight of each link within a CS, and it is calculated as
the average PLV across the edges that make up the clique of CS. This Cluster
PLV represents the strength of the synchronization within each CS and
defines the hierarchy of CS according to its strength.

The use of PLV is widespread in the literature on synchronization, in
particular related to the Kuramoto model. The purpose is to remove the
amplitude effects and model the level of synchronization as just the
average across time of phase difference between signals as an
estimation of their Kuramoto order parameter
\cite{ponce2015resting}. As discussed in the Results, local fibration symmetries and cluster synchronization patterns are in a
correspondence, so a coherent implementation of the fibration symmetry
framework needs a measure of synchronization to start with. Coupled oscillators have been widely used to describe
large-scale brain activity
\cite{deco2017dynamics,menara2022functional}, and the Kuramoto local
order parameter represents a spatial average of the complex phase
factor of the local oscillators weighted by the coupling
\cite{strogatz2000kuramoto}, offering a robust measure of
synchronization between fluctuating signals.

Moreover, recent findings \cite{pang2019power, gravel2018phase, bolt2018inter, kringelbach2020brain, honari2021evaluating, honari2022mode,
  xu2020inter} support our choice of making use of phase
synchronization to analyze the BOLD signal, even if its low-frequency
content (typically 0.01--0.1 Hz) does not guarantee a pure oscillatory process
behind its intrinsic fluctuations. 

Nonetheless, we repeated the analysis using the Pearson correlation matrix instead of the PLV, since correlation between BOLD signals is a common measure of their coupling. With the group-averaged resting-state correlation matrix, the same four clusters satisfy Eq.~(\ref{eq:1}) (\SuppFig{extended_figure_12}).

\section{Cluster synchronization in resting state and tasks}
\label{sec:restingtask}

\begin{figure*}[t!]
\centering
\includegraphics[width=\textwidth]{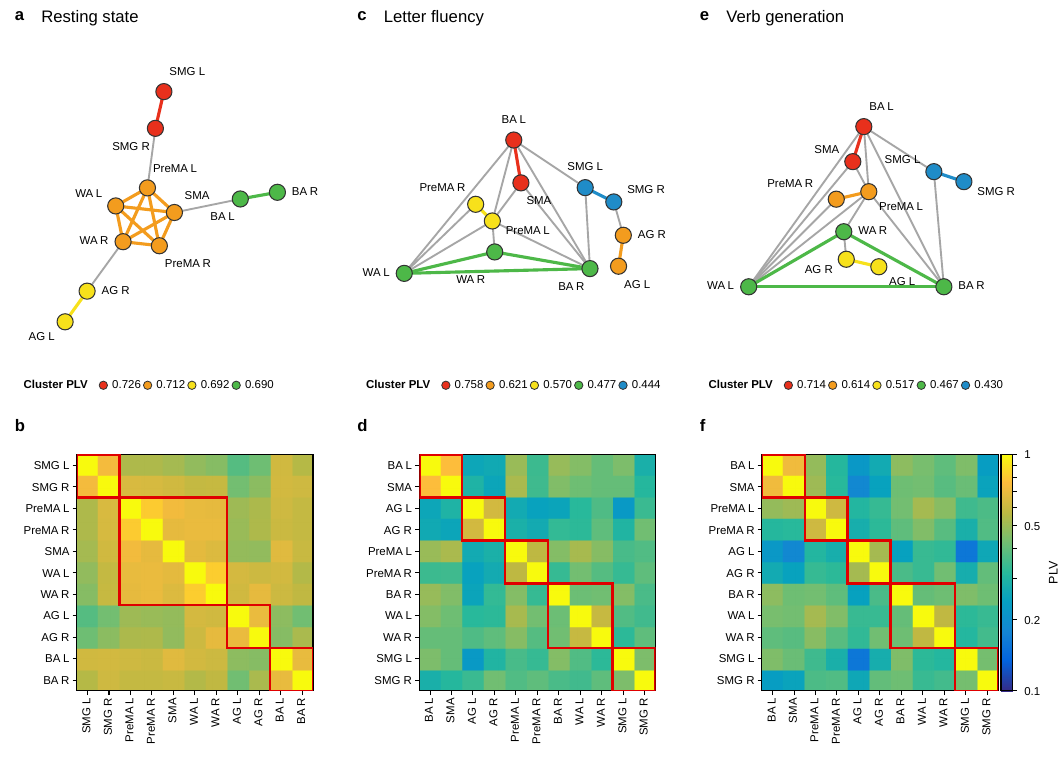}
\caption{{\bf Functional networks and cluster synchronization.} {\bf a} Functional network in the resting state (RS); cluster synchronization (CS) cliques are shown in different colors. {\bf b} Group-average phase locking value (PLV) matrix of the eleven regions of interest (ROIs) in RS; the PLV is corrected against surrogate data (Eq.~(\ref{eq:plv_corrected}), Methods). {\bf c} Functional network during the letter fluency task. {\bf d} PLV matrix during the letter fluency task. {\bf e} Functional network during the verb generation task. {\bf f} PLV matrix during the verb generation task. In the networks in {\bf a}, {\bf c} and {\bf e}, nodes and links of each CS are colored by the rank of its Cluster PLV, the average PLV over the links of the CS (values in the key, from the highest, red, to the lowest); grey links connect different CS and are the links obtained by the percolation procedure (links added in decreasing order of PLV) down to the PLV at which all ROIs are connected. The red boxes in {\bf b}, {\bf d} and {\bf f} indicate the CS, shown in decreasing order of Cluster PLV. Matrices are averages over $n=20$ participants. ROI abbreviations: SMA, supplementary motor area; PreMA, premotor area; SMG, supramarginal gyrus; BA, Broca's area; AG, angular gyrus; WA, Wernicke's area; L and R, left and right hemisphere. Source data are provided as a Source Data file.}
\label{fig:functional}
\end{figure*}

\begin{figure*}[t!]
\centering
\includegraphics[width=\textwidth]{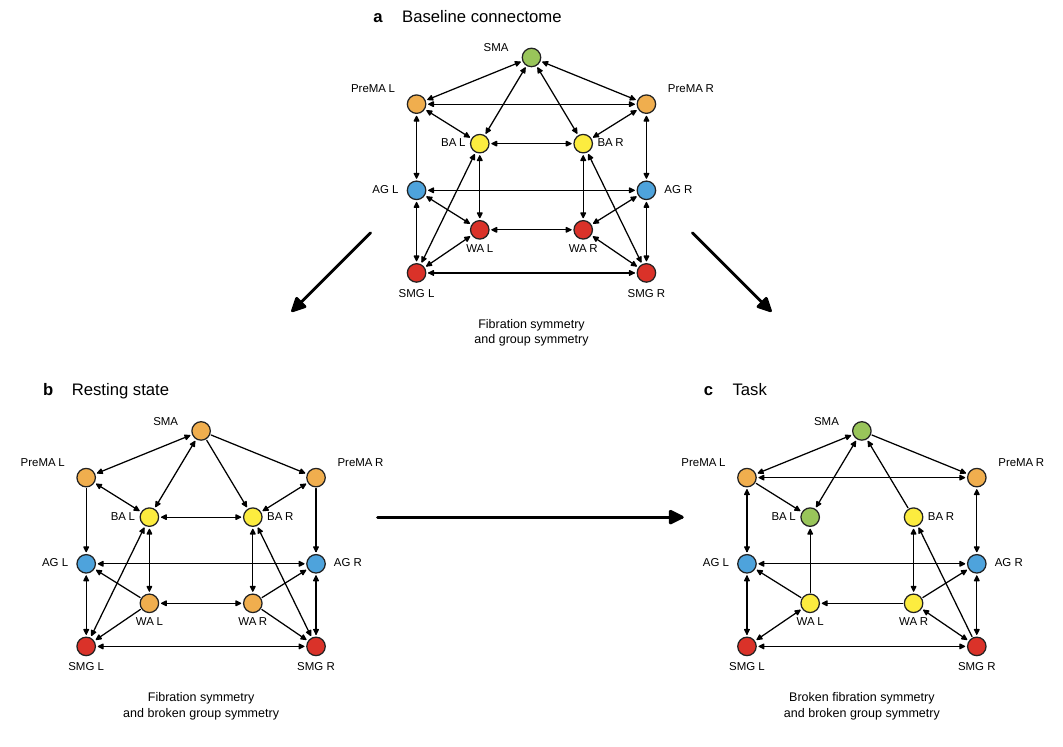}
\caption{{\bf Symmetry breaking from the baseline connectome to the inferred resting-state and task networks.} {\bf a} Minimal balanced coloring of the baseline connectome. This network has the highest group symmetry: a global automorphism group that coincides with the local fibration symmetry, with five orbits equal to five fibers (five balanced colors). {\bf b} Structural network inferred for the resting state (RS) from the cluster synchronization (CS) of Fig.~\ref{fig:functional}a. The network has only local fibration symmetry, with four fibers, and no global symmetry, which is broken with respect to the connectome in {\bf a}. {\bf c} Structural network inferred for the language tasks from the coloring of Fig.~\ref{fig:functional}c or Fig.~\ref{fig:functional}e (which are identical). The lateralization of function under the language task breaks the fibration symmetry of {\bf b}, yielding more fibers than in RS; the group symmetry remains broken. ROI abbreviations: SMA, supplementary motor area; PreMA, premotor area; SMG, supramarginal gyrus; BA, Broca's area; AG, angular gyrus; WA, Wernicke's area; L and R, left and right hemisphere. The networks in {\bf b} and {\bf c} are, among the configurations with the minimum number of removed links, those selected by the percolation order of the tract weights (Methods Sect.~\ref{sec:algorithm}). Source data are provided as a Source Data file.}
\label{fig:broken}
\end{figure*}

The patterns of CS found within the language network allow
discriminating the resting state condition from the task ones. RS-fMRI
demonstrates sub-optimal characterization of both language dominance
and lateralization of eloquent areas \cite{teghipco2016disrupted}, due to enhanced homotopic synchronization. This is especially true in
networks with left-right symmetry, such as those involved in motor and
vision, as well as in language, which is normally lateralized
(breaking the left-right symmetry) during the execution of the task
\cite{seitzman2019state}. Our results confirm this evidence by
demonstrating high left-right symmetry of the language network during
resting-state. We find, according to a descending synchronization hierarchy, a CS composed of bilateral SMG, a pentagonal CS composed of two bilateral pairs of regions (PreMA
and WA) and the SMA, a CS composed of bilateral AG, and a CS composed of bilateral
BA (Fig.~\ref{fig:functional}a).

The clusters that we find in RS are hierarchically ordered according
to the Cluster PLVs as follows (Fig.\ref{fig:functional} a and b) :
\noindent the pair \{SMG L, SMG R\} [PLV = 0.726], the pentagon \{PreMA L, PreMA R, SMA, WA L, WA R\} [PLV = 0.712], the pair \{AG L, AG R\} [PLV = 0.692] and the pair \{BA L, BA R\} [PLV = 0.690].  The inter-cliques connections were characterized by PLV
values smaller than the Cluster PLVs: (AG R,
WA R) with [PLV = 0.682], (BA L, SMA) with [PLV = 0.662]
and (SMG R, PreMA L) with [PLV = 0.639].

The largest synchronization clique composed of SMA, PreMA
(bilateral), and WA (bilateral) fits the auditory-motor integration
mechanism of the dorsal stream of the language processing model.  As a
consequence of the internal forward model, the clique can act as a
motor speech unit that, once activated, predicts auditory consequences
that can be checked against the auditory target. If they match, that
unit will continue to be activated, resulting in an articulation that
will hit the target. If there is a mismatch, a correction signal can
be generated to activate the correct motor unit. The predictions are
assumed to be generated by an internal model that receives efference copies of motor commands and integrates them with information about
the current state of the system and experience (learning) of the
relation between particular motor commands and their sensory
consequences \cite{hickok2011sensorimotor}. The resting state
largest synchronization clique enhances the preparatory
configuration for the auditory-motor integration to efficiently run
when single regions are engaged. The five-node cluster is also found with time series averaged within Glasser parcels in the whole-brain dataset of the same participants (Methods Sect.~\ref{sec:pentagon-direct}).

The RS functional network undergoes changes when a task is performed.
The letter fluency task (Fig.\ref{fig:functional}c and \ref{fig:functional}d) and the
verbs generation task (Fig.\ref{fig:functional}e and \ref{fig:functional}f) returned
very similar patterns of synchronization, both showing the clique
formed by SMA and BA L as the most synchronized one [PLV = 0.758 and
  PLV = 0.714, respectively]. The other cliques have Cluster PLV 0.621 (AG L, AG R), 0.570 (PreMA L, PreMA R), 0.477 (WA L, WA R, BA R) and 0.444 (SMG L, SMG R) in the letter fluency task, and 0.614 (PreMA L, PreMA R), 0.517 (AG L, AG R), 0.467 (WA L, WA R, BA R) and 0.430 (SMG L, SMG R) in the verb generation task.

Covert letter fluency tasks require phonologic access, verbal
working memory, and lexical search activity, which grant a strong
activation and lateralization of frontal areas \cite{li2017lexical,
  corrivetti2019dissociating}. Sentence completion and verb
generation tasks require word recognition and comprehension,
understanding of syntactic–semantic relationships between words,
planning of a sentence structure and word retrieval
\cite{li2017lexical}. This leads to increased recruitment of
temporoparietal language-related areas, including WA, SMG, and AG
\cite{li2017lexical,unadkat2019functional}.

We acknowledge that the hemodynamic response function (HRF) differs across the whole brain, inducing spatial phase lags and spurious synchronization/correlation measures of the BOLD signal. However, brain regions functionally relevant for the execution of specific tasks are characterized by a minimal distribution of phase lags and spatial amplitudes\cite{taylor2018characterization}, making the correlation/synchronization of their BOLD timeseries minimally affected by such bias.

We tested this conclusion by analyzing resting state data after hemodynamic response function estimation and deconvolution \cite{wu2021rshrf}. \SuppFig{extended_figure_deconvolved} and \SuppFig{extended_figure_deconvolved_glasser} show the mean PLV matrix calculated from time series deconvolved by HRF and averaged within spheres and Glasser parcels respectively. In the deconvolved data with spherical ROIs, the same four clusters satisfy Eq.~(\ref{eq:1}). With Glasser parcels, the four clusters (PF, five-node, PFm and area 44) also satisfy Eq.~(\ref{eq:1}) within the PLV tolerance of the control analyses (Sect.~\ref{sec:wholebrain}).

To verify that the CS of the tasks are not affected by the frequency content of the signals, we repeated the analysis after band-pass filtering the task time series in the same frequency range used for the resting state (0.01--0.1 Hz), so that rest and tasks are compared within the same frequency band. In both tasks, the same CS cliques satisfy Eq.~(\ref{eq:1}) (\SuppFig{bandpass}): SMA and left BA, bilateral PreMA, bilateral AG, right BA with bilateral WA, and bilateral SMG.

\subsection{Whole-brain analysis and control for shared/external input.}
\label{sec:wholebrain}

A test of whether each bilateral pair is more synchronized with its partner than with any other cortical region is not informative for this purpose. The BOLD signal carries strong short-range spatial autocorrelation (shared vasculature, partial-volume and smoothing effects), so that adjacent cortical parcels are, for essentially any seed region, more synchronized with one another than functionally related but spatially separated regions. We verified this directly: when each of the bilateral pairs reported here was compared against the 349 cortical parcels outside the 11 ROIs, every pair was exceeded by an adjacent parcel in the same hemisphere (for example, SMG by its neighbor, the perisylvian language area PSL, and BA by its neighbor 6r). This is a generic property of cortical BOLD data rather than a specific feature of the language network or of our ROI selection, and an unconstrained whole-cortex comparison is therefore not a meaningful test for any predefined ROI set. Instead, we tested whether the reported clusters, evaluated within the hypothesis-driven 11-ROI set (Methods Sect.~\ref{sec:dual}), survive the explicit removal of specific, literature-motivated candidate sources of shared signal.

To address whether the reported functional clusters could be explained by
shared or externally-driven input common to many brain regions, rather than by
network-intrinsic synchronization, we obtained whole-brain resting-state fMRI
time series from all 20 participants, registered to the same Human Connectome Project multimodal parcellation (HCP-MMP1.0; Glasser atlas) used for the 11-ROI language network, plus a
subcortically-extended atlas comprising bilateral thalamus, the thalamic
auditory relay (medial geniculate nucleus, MGN), and the inferior colliculus
(IC), pre-resampled to the resting-state functional grid. From the 360
cortical parcels we derived three candidate shared-signal regressors: the
global cortical signal (mean over all 360 parcels), the mean signal in
bilateral primary auditory cortex (A1), and the mean signal in bilateral
perisylvian language area (PSL). Each candidate regressor (and the three
subcortical structures) was in turn removed from the 11 ROI time series by
ordinary least-squares regression (design matrix with the candidate signal
plus an intercept; the fitted intercept was retained so as to remove only
confound-explained variance), the surrogate-corrected PLV matrix (Eq.~(\ref{eq:plv_corrected})) was recomputed from the residualized time series and the CS partition reported for the resting state was tested with Eq.~(\ref{eq:1}). In this and in the other control analyses (Sect.~\ref{sec:task-generalization} and the analyses with different parcellation, deconvolution or filtering), differences in PLV smaller than 0.01, well below the standard error of the group-average PLV across participants (about 0.05), were not considered meaningful: a clique was taken to satisfy Eq.~(\ref{eq:1}) when its Cluster PLV was less than 0.01 below the strongest link to an ROI not yet assigned. Nine conditions were tested in total: raw; global signal alone; A1
alone; PSL alone; thalamus alone; MGN alone; IC alone; global+A1+PSL
combined; and all six candidate signals combined.

Table~\ref{tab:wholebrain-rs} reports (see also Fig.~\ref{fig:wholebrain_heatmap}), for each condition, the Cluster PLV of the four resting-state clusters (SMG-SMG, the five-node cluster SMA-PreMA-PreMA-WA-WA, AG-AG and BA-BA), together with whether each satisfies Eq.~(\ref{eq:1}) under that condition. All four clusters satisfy Eq.~(\ref{eq:1}) in all nine conditions. The removal of cortical signals reduces the Cluster PLVs most: the global signal mainly affects the five-node cluster (from 0.697 to 0.250), and the PSL signal, from the region adjacent to SMG, mainly affects SMG-SMG (from 0.746 to 0.338). Bilateral thalamus, MGN and IC, regressed individually, reduce the Cluster PLVs by at most 0.13 and do not change the validity of any cluster, indicating that the shared component of the observed pattern is cortical rather than subcortical; the effect of PSL on SMG-SMG is consistent with the spatial proximity of the two regions and with the short-range autocorrelation of the BOLD signal discussed above.

\subsection{Five-node cluster in the whole-brain dataset.}
\label{sec:pentagon-direct}

In the whole-brain resting-state dataset ($n=20$), in which ROI time series are averaged within Glasser parcels, the five-node cluster (SMA, PreMA$_L$, PreMA$_R$, WA$_L$, WA$_R$) has Cluster PLV 0.697, close to the value reported in the main text (0.712), and satisfies Eq.~(\ref{eq:1}) together with the other three resting-state clusters (Table~\ref{tab:wholebrain-rs}).

\subsection{Task-state generalization: letter fluency and verb generation.}
\label{sec:task-generalization}

To test whether the whole-brain analysis and external-input control
above generalize beyond the resting state, we repeated the identical
analysis on the two task-fMRI datasets of the same 20
participants: a letter-fluency task and a verb-generation task (Methods
Sect.~\ref{sec:languagetask} for the task protocols). Both tasks use an
identical block design (8 epochs $\times$ 50~s; 20~s task + 30~s rest).
Functional volumes were registered to the same HCP-MMP1.0 (Glasser)
parcellation and subcortical atlas used above, and mean time series were
extracted for all 360 cortical parcels, the 11 language ROIs, and the three
subcortical structures for every participant and both tasks.

Because the experimental task is itself the external field whose relationship to the reported network structure is of interest, rather than a nuisance signal to be removed, the raw (task-response not
regressed out) time series were used as the primary analysis for both
tasks, on the explicit rationale that regressing out the task design would
remove exactly the lateralizing/symmetry-breaking signal the task-state
hypothesis concerns. The surrogate-corrected PLV was computed
exactly as in Sect.~\ref{sec:wholebrain}, the task CS partition (SMA-BA$_L$, PreMA-PreMA, AG-AG, BA$_R$-WA-WA and SMG-SMG; Fig.~\ref{fig:functional}c-f) was tested with Eq.~(\ref{eq:1}), and the identical nine-condition
whole-brain regression battery was applied independently to each task (Tables~\ref{tab:suppl-lf} and \ref{tab:suppl-vg}).

The dominant cluster in both task datasets is the same fronto-opercular pairing highlighted by the main-text task-state analysis
(SMA-BA$_L$; Sect.~\ref{sec:restingtask}; Fig.~\ref{fig:functional}c-f): it
has the highest Cluster PLV in raw data for both tasks (letter
fluency 0.789; verb generation 0.827) and satisfies Eq.~(\ref{eq:1}) in all nine control
conditions, including the all-signals-combined
condition (letter fluency 0.552; verb generation 0.608). In both tasks, all five task clusters satisfy Eq.~(\ref{eq:1}) in every condition. As in the resting state, the PSL signal mainly affects SMG-SMG (letter fluency from 0.659 to 0.370; verb generation from 0.597 to 0.303).

Taken together, this whole-brain analysis and external-input control, covering resting state and two task paradigms in the same 20-participant cohort, shows that the CS partitions reported in the main text are robust to shared/external-input
regression, with SMG-SMG showing a consistent, anatomically motivated
dependence on perisylvian-language (PSL) signal across all three datasets.

\subsection{Dynamical validation via Kuramoto oscillator simulation.}
\label{sec:kuramoto}

To provide direct dynamical support for the claim that the reported
fiber partitions correspond to genuine, flow-invariant synchronization
states rather than incidental graph colorings (ref.~\cite{deville2015modular}, Theorem 4.3.1), we simulated directed Kuramoto oscillator dynamics
$\dot\theta_i = \omega_i + K\sum_j A_{ij}\sin(\theta_j-\theta_i)$ on
structural subgraphs with known fiber partitions, assigning a common
natural frequency and a common initial phase to the nodes of the same analytic
fiber (as required for the flow-invariant cluster-synchronization
subspace) and different frequencies across fibers (equally spaced in $[-3,3]$ and randomly assigned); the initial phase of each node was then perturbed around the common phase of its fiber (Gaussian noise, standard deviation 0.6 rad) so that the return to within-fiber synchrony tests its stability. From the simulated
phase trajectories (second half of a run of 3,000 time units) we computed the PLV matrix, grouped the oscillators connected by links with PLV $\geq 0.9$,
and compared the dynamically observed clusters with the target fiber partition, over 500 random draws of frequencies and initial phases. We used a weak coupling, $K=0.3$, so that oscillators of different fibers, driven at different natural frequencies, remain distinguishable.

We first verified that the 11-ROI baseline connectome (Methods
Sect.~\ref{sec:dual}) computationally reproduces the reported symmetry
structure exactly: 5 orbits (automorphism group, computed via exact VF2
isomorphism search) and 5 fibers (balanced coloring via color refinement),
with fibers equal to orbits, confirming the high-symmetry condition
reported in the main text.

\begin{figure*}[t!]
\centering
\includegraphics[width=0.5\textwidth]{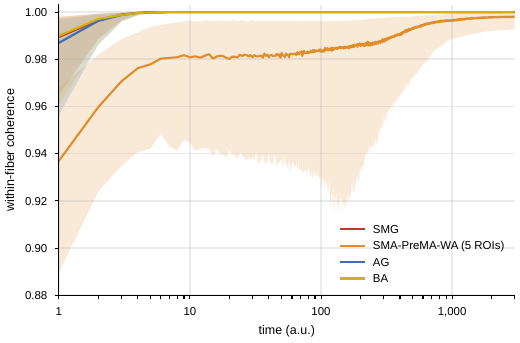}
\caption{{\bf Dynamical validation of the structure-function-structure loop.} Within-fiber phase coherence over time (logarithmic axis) in Kuramoto oscillator simulations (coupling strength $K=0.3$, runs of 3,000 time units) on the resting-state subgraph inferred by the mixed integer linear programming (MILP) inference (Fig.~\ref{fig:broken}b), with natural frequencies equal within fibers and initial phases perturbed within each fiber. Lines show the median and shaded areas the interquartile range across 500 random draws of frequencies and initial phases (Methods Sect.~\ref{sec:kuramoto}). ROI abbreviations: SMA, supplementary motor area; PreMA, premotor area; SMG, supramarginal gyrus; BA, Broca's area; AG, angular gyrus; WA, Wernicke's area; L and R, left and right hemisphere. Source data are provided as a Source Data file.}
\label{fig:kuramoto}
\end{figure*}

We then used the resting-state fiber partition reported in the Results (Fiber$_1$ = \{SMA, PreMA$_L$, PreMA$_R$,
WA$_L$, WA$_R$\}, Fiber$_2$ = \{BA$_L$,BA$_R$\}, Fiber$_3$ =
\{AG$_L$,AG$_R$\}, Fiber$_4$ = \{SMG$_L$,SMG$_R$\}; Methods
Sect.~\ref{sec:restingtask}) as the MILP target; this partition differs from the native fiber partition of the baseline and requires non-trivial edge removal to be recovered. The deletion-only MILP removed 10 of
46 directed edges (22\%) and recovered a subgraph whose analytic fiber
partition matches the RS target exactly (Fig.~\ref{fig:broken}b). Simulating Kuramoto dynamics on
this subgraph, the BA, AG and SMG fibers reached full within-fiber synchrony in all draws, and the PLV clusters reproduced the RS partition in 489 of 500 draws. In the other 11 the five-node fiber was still split into \{SMA, PreMA$_L$, PreMA$_R$\} and \{WA$_L$, WA$_R$\}, which are not directly connected within the fiber and phase-lock only after long transients. The five-node fiber phase-locked, but not always with identical phases (within-fiber coherence above 0.99 in 440 draws). The simulated dynamics thus close the structure~$\to$~function~$\to$~structure loop (Fig.~\ref{fig:kuramoto}).

\subsection{Weighted extension of the balanced-coloring constraint and quasi-fibration analysis.}
\label{sec:weighted}

In a weighted network, the balanced-coloring condition depends on the total incoming connection strength rather than on the in-degree. We therefore generalized the deletion-only MILP's
constraint from binary in-degree equality to weighted in-strength
equality: for every pair of nodes assigned the same color and every source
color class, the retained edges' summed weight from that source class into
each node must be equal (or, in the tolerance-band version below, equal to
within an explicit slack $\varepsilon$ for every pair of nodes of the same color). Feasibility of this weighted
constraint was tested for three target partitions: the baseline
connectome's own native 5-fiber partition, the resting-state 4-fiber
partition, and the task fiber partition (5 fibers;
Methods Sect.~\ref{sec:restingtask}).

Under exact weighted equality ($\varepsilon=0$) and unweighted (all tract
weights equal to 1) edges, all three partitions are exactly feasible, as
expected since this reduces to the binary formulation used in the Results. We then introduced a single, literature-grounded synthetic
asymmetry, namely a leftward volumetric asymmetry factor $\lambda$ applied to the left arcuate fasciculus tract weight (BA$_L$-WA$_L$) relative to its
right-hemisphere homolog (BA$_R$-WA$_R$), reflecting the documented
leftward volumetric asymmetry of this tract in the general population, and evaluated the imbalance this introduces into each partition's exactly-balanced ($\lambda=1$) structural solution. This
imbalance equals $|\lambda-1|$ for all three partitions alike: in each solution the perturbed tract changes by $|\lambda-1|$ the input of a node relative to another node of the same fiber (left versus right BA in the baseline and resting-state solutions, left BA versus SMA in the task solution), so none of the three solutions is
selectively immune to this specific single-tract perturbation, and a
tolerance $\varepsilon$ equal to the fractional asymmetry itself is
sufficient to restore feasibility of each partition's own known solution.
This characterizes the sensitivity of each partition's own known solution to this single-tract perturbation.

\begin{figure*}[t!]
\centering
\includegraphics[width=0.5\textwidth]{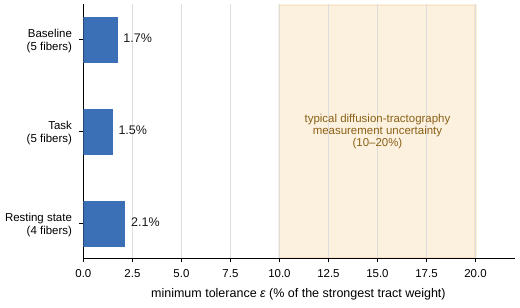}
\caption{{\bf Weighted balanced-coloring feasibility under empirical structural weights.} Minimum tolerance $\varepsilon$, expressed as a percentage of the strongest single tract weight, required for the feasibility of the weighted balanced-coloring constraint for each of the three target partitions (baseline connectome, resting state and task), under diffusion-tractography weights from Rosen and Halgren \cite{rosen2021whole} ($N=1065$ Human Connectome Project participants). All three values fall below the typical range of diffusion-tractography measurement uncertainty (shaded area) \cite{jones2010precision}. Source data are provided as a Source Data file.}
\label{fig:weighted_tolerance}
\end{figure*}

A further test used real, empirically-measured structural weights
rather than a synthetic single-tract perturbation: the whole-cortex
probabilistic diffusion-tractography connectome of Rosen \& Halgren
\cite{rosen2021whole} (N~=~1065, HCP subjects, registered to the same
HCP-MMP1.0/Glasser parcellation used throughout this work), from which we extracted the $11\times11$ submatrix corresponding to the 11 ROIs (linear tract weight $=10^{F_{\mathrm{pt}}}$, where $F_{\mathrm{pt}}$
is the log-scaled fractional streamline probability reported by Rosen \&
Halgren). Under these real weights, none of the three partitions is
exactly balanced (as expected of continuously-valued empirical weights),
but all three require only a small tolerance, expressed as a percentage of
the single strongest tract weight in the submatrix, to become feasible:
baseline 1.7\%, task 1.5\%, and resting state 2.1\% (Fig.~\ref{fig:weighted_tolerance}). All three values are well within reported
diffusion-tractography test-retest and inter-algorithm measurement
uncertainty, commonly reported in the 10--20\% range for comparable
dorsal-stream tracts \cite{jones2010precision}. We interpret this as
evidence for approximate, rather than exact, weighted fibration symmetry
(a quasi-fibration in the sense of \cite{boldi2022quasifibrations}); the
resting-state partition, which is the most symmetric of the three
(4~fibers vs.\ 5), is correspondingly the most sensitive to residual
weight heterogeneity in the real-data test.

As a final, cautionary methodological note, we also tested a
heavily-thresholded, publicly available group-consensus structural
connectome (Schaefer-800 parcellation, processed as described by Park et al.\ \cite{park2021signal} and released by Bazinet et al.\ \cite{bazinet2023assortative},
obtained from the netneurolab/bazinet\_assortativity repository), matching
our 11 ROIs to the nearest parcel by standard literature MNI coordinates
(not the Glasser parcellation or the authors' own ROI masks, so this test
is illustrative rather than a direct substitute for the Rosen \& Halgren
analysis above). At the 3\% edge density this public connectome retains
after its own thresholding pipeline, all five bilateral homotopic callosal
connections among our 11 ROIs, including the strong, well-established left-right connection of Broca's area, have exactly zero weight,
underscoring that not every published, thresholded group connectome is
suitable for testing hypotheses about these specific tracts, and
which motivates the literature-curated baseline connectome and the
Rosen \& Halgren whole-cortex probabilistic connectome used elsewhere in
this Methods section.

\section{Definition of automorphisms and group symmetries}
\label{sec:automorphism}

Basic graph theoretical definitions \cite{dixon1996permutation}:

\begin{definition}[{\bf Graph}] 
A graph $G=(N_G,E_G)$ is defined by a set $N_G$ of nodes and a set $E_G$ of arcs, endowed with two functions $s,t: E_G \to N_G$ that associate 
a source and target node with each edge. 
\end{definition}

\begin{definition} [{\bf Permutation of a graph}].
A permutation $\pi$ of a graph $G(N_G,E_G)$ is a bijective map
from the set of nodes $N_G=\{1,\dots,N\}$ to itself, $\pi:N_G\to N_G$,
that represents the permutation of the node labels.
\label{permutation}
\end{definition}

For example, the permutation $\pi$ in the graph of the dual stream
baseline connectome in Fig. \ref{fig:intro}a that maps BA$_L$ to
BA$_R$ while leaving all the other nodes alone is denoted in
   cycle notation: 
   
   \begin{equation} 
   \pi = ({\rm BA}_L\,{\rm BA}_R) ,
   \end{equation}
   meaning that BA$_L \to$ BA$_R \to$ BA$_L$. 

\begin{definition}[{\bf Permutation matrix}]
A permutation matrix $P$ is an $N \times N$ matrix that is
obtained from the identity matrix by permuting both rows and columns
according to $\pi$.
\end{definition}

For a graph with $N$ nodes, there are $N!$ permutations, some of which
are permutation symmetries or automorphisms, and the rest are not.  The
set of all permutations of the nodes of a graph $G(N_G,E_G)$ forms a
group $\mathbb{S}_N = \{P_1, \dots P_K\}$ where $K = N!$. It is called
the symmetric group (not to be confused with a symmetry group).
This set forms a group because the permutations satisfy the
associative property, composition (composing two permutations leads to
another permutation), and have an identity and inverse.

Basic group theoretical definitions \cite{aschbacher2000finite}:

\begin{definition}[{\bf Graph homomorphism}]
A graph homomorphism $\varphi: G \to H$ (from a graph $G$ to a
graph $H$) is a pair of functions $\varphi_N: N_G \to N_H$ and
$\varphi_E: E_G \to E_H$ such that $s(\varphi_E(a))=\varphi_N(s(a))$
and $t(\varphi_E(a))=\varphi_N(t(a))$ for every edge $a \in E_G$.
\end{definition}

\begin{definition}[{\bf Graph isomorphism}]
A graph isomorphism $\varphi: G \to H$ is a graph homomorphism
whose components $\varphi_N$ and $\varphi_E$ are both bijections.
\end{definition}

\begin{definition}[{\bf Graph Automorphism}]

A (graph) automorphism (also called a symmetry permutation or
group symmetry) $\pi: G \to G$ is an isomorphism from a graph to
itself.

Alternatively, an automorphism is a permutation $\pi : G \to G$ of the vertex set $N_G$, such that the pair of vertices $i$ and $j$ forms an
edge $(i,j)$ if and only if $(\pi(i), \pi(j))$ also forms a edge.
\end{definition}

That is, an automorphism preserves adjacency and non-adjacency of all the nodes in the graph, and therefore, it is a global symmetry: two nodes are adjacent after the permutation if and only if they were adjacent before the permutation.

Any permutation matrix of a permutation $\pi$
transforms the adjacency matrix into another $A'$ as $A' = P A
P^{-1}$. If $P$ represents an automorphism, then $A'=A$, and
\begin{equation}
\label{E:perm_auto}
    A = P A P^{-1}.
\end{equation}

Since the group consists of matrices, we can state this condition
differently. Equation \eqref{E:perm_auto} holds if and only if
the matrix $P$ commutes with $A$, so $PA=AP$. Equivalently, 
the commutator is zero:

\begin{equation}
  [P, A] = P A - A P = \textbf{0} .
\end{equation}

\begin{definition}[{\bf Symmetry group}]
Graph automorphisms form a group with respect to function composition; this group is denoted by $\mathrm{Aut}(G)$:
\begin{equation}
\mathrm{Aut}(G) = \{\pi \,|\, \pi \text{ is a symmetry permutation of } G \} .
\end{equation}
\end{definition}
The group   $\mathrm{Aut}(G)$ acts  on the set $N_G$, mapping the pair $(\pi,x) \in \mathrm{Aut}(G) \times N_G$ to $\pi(x) \in N_G$. The order of a finite group is the number of its elements.  The generators of the symmetry group are a subset of the group set such that every element of the group can be expressed as a composition of finitely many elements of the subset and their inverses.

The set of graph automorphisms permutes certain subsets of nodes
among each other.  When the symmetry group ${\rm Aut}(G)$ acts on the
network, a given node $i$ is moved by the permutations of the group to
various other nodes $j$.  In the language of group theory, the set of
all nodes to which $i$ can be moved defines the orbit of
node $i$, which in turn defines the orbital partition of the network.

\begin{definition}[{\bf Orbit of a node}]
The orbit of a node $i \in N_G$ for the symmetry group
${\rm Aut}(G)$ is:
    \begin{equation}
        \mathcal{S}(i) = \{j \in N_G \,|\, \exists \, \pi \in {\rm Aut}(G) \text{ such that } \pi(i) = j \}.
    \end{equation}
    \label{orbits}
    \end{definition}
    It can easily be proved that two orbits $\mathcal{S}(i)$ and
    $\mathcal{S}(j)$ are equal or disjoint, and the union of
    all orbits equals $N_G$.  Therefore, the set of all orbits
    induces a partition of the nodes into mutually disjoint clusters.
    This set of all orbits forms the orbital partition.  The
    same definition can be applied to subgroups $H$ of ${\rm Aut}(G)$
    to obtain $H$-orbital partitions. When $H={\rm Aut}(G)$, we obtain
    the partition into the fewest subsets.

The orbital partition of the symmetry group corresponds to clusters of
nodes that synchronize under a suitable dynamical system of equations
that is admissible to the graph. In other words, the orbits guarantee
that the cluster synchronization subspace is flow-invariant
\cite{pecora2014cluster,golubitsky2006nonlinear}.

The orbits of the symmetry group (i.e., the partition of $N_G$ into
orbital equivalence classes, where $x$ is equivalent to $y$ if and
only if $\pi(x)=y$ for some automorphism $\pi$) define the
automorphism symmetry of $G$.

\section{Definition of fibration symmetries}
\label{sec:fibration}

\begin{definition}[{\bf Graph Fibration}]
A homomorphism $\phi: G \to B$ is a fibration if and
only if for every $a \in E_B$ and every $x \in N_G$ such that
$\phi(x)=t(a)$, there exists exactly one $a' \in E_G$ such that
$t(a')=x$ and $\phi(a')=a$. This unique arc $a'$ is called the
lifting of $a$ at $x$ \cite{boldi2002fibrations}.
\end{definition}

\begin{definition}[{\bf Fibers of the  Fibration}]
The fibers of $\phi$ are the sets of nodes of $G$ that are mapped to the same node of $B$: these sets form the fiber partition of $N_G$.
\end{definition}

\begin{definition}[{\bf Input tree of a node}]
Given a graph $G$ and a node $x \in N_G$ the input tree of $x$ in $G$, $T_G(x)$, is defined recursively as follows: it is a (typically, infinite) rooted tree whose root has as many children as there are in-neighbors of $x$ in $G$, and such that the subtrees rooted at each child is the input tree of the corresponding in-neighbor in $G$. 
\end{definition}

It is easy to see that if $x,y$ are two nodes of $G$ that are in the
same fiber of some fibration, then $T_G(x)$ and $T_G(y)$ are
isomorphic trees.

\begin{definition}[{\bf Symmetry  Fibration}]
For every graph $G$, there exists a (base) graph $B$ and a surjective
fibration $\mu: G \to B$ such that two nodes are in the same fiber of
$\mu$ if and only if $T_G(x)$ and $T_G(y)$ are isomorphic. This
surjective fibration is essentially
unique~\cite{boldi2002fibrations}. It collects all the symmetries of
the graph and produces the (minimal) base with the minimal number of
fibers: it is called the symmetry fibration
\cite{morone2020fibration}, and its fibers define the fibration
symmetry of $G$.
\end{definition}

\begin{definition} [{\bf Cluster synchronization (CS) in a fiber}]
Cluster synchronization occurs for all nodes in a fiber, and
they have the same dynamic state as node $i$, i.e.,
\begin{equation}
  x_i(t)=x_j(t) \,\,\,\, \mbox{ if}\,\,\,\, j \in {\mathrm{Fiber}(i)} ,
\end{equation}
\end{definition}
This cluster is nontrivial only for fibers of size $> 1$.

\begin{definition} [{\bf Minimal base of the symmetry fibration}]
Collapsing the nodes in each fiber of the minimal fiber partition to
form a single node and respecting the lifting property provides the
minimal base of the symmetry fibration.
\end{definition}

The partition of nodes into fibers of a fibration coincides with the
partition obtained by balanced coloring, or equitable partition
~\cite{golubitsky2006nonlinear,stewart2003symmetry,boldi2002fibrations}. This
correspondence provides a third alternative definition of a graph
fibration in terms of the input sets rather than the input trees:

 A balanced coloring of a graph is an assignment of colors to
  nodes, such that nodes of one color receive the same amount of
  each color from their in-neighbors (i.e., the same number of
  in-neighbors of each color, see Fig. \ref{fig:symmetry}b, right).

Aldis~\cite{aldis2008polynomial} [Theorem 4.2 and Corollary 4.3] has
indeed shown that the fibers of the fibration are the partition
induced by balanced colorings of the graph.  Thus, we identify the CS
obtained from the dynamics with the fibers of the graph or analogously
the balanced coloring.

The orbital partition obtained from automorphisms
(Fig. \ref{fig:symmetry}a, right) is also a balanced coloring but is
generally finer than the coarsest balanced coloring determined by the
symmetry fibration: i.e., every orbit is contained in a fiber, but not every fiber
is an orbit. This implies that a graph may have more fibration symmetries than those induced by the automorphisms (Fig.
\ref{fig:symmetry}b).

In other words, all automorphisms are fibrations, but not all fibrations are automorphisms. Intuitively, the conditions imposed by
automorphisms, being on non-local scales, are much harder to satisfy
than in fibrations, which preserve only the local
in-neighborhood. Algorithms exist to efficiently calculate the minimal
balanced colorings (fibers) in a network 
\cite{kamei2013computation,morone2020fibration,leifer2020circuits}. They are also widely used in machine learning and in graph neural networks (GNNs) as the
Weisfeiler-Lehman graph isomorphism test \cite{xu2018powerful}.

It must be noted that the study of shared input profiles in brain networks has previously been addressed through the matching index (MI), a heuristic similarity measure between pairs of nodes based on their common neighbors \cite{song2014spatial,chen2017features}. Although MI-based approaches have proven useful for predicting connectivity and identifying functional clusters, the fibration framework differs in fundamental ways. First, fibration symmetry is an exact algebraic condition grounded in category theory, with a rigorous dynamical guarantee: nodes in the same fiber synchronize under any admissible dynamics \cite{deville2015modular}, a property that MI does not provide. Second, fibrations identify the global partition of the network into equivalence classes (fibers) satisfying a balanced-coloring condition, rather than computing pairwise similarity scores. These imply that the fibration framework is not simply a symmetry detection method per se, but it is able to characterize symmetry-breaking transitions accompanying functional state changes, a feature that MI cannot replicate.

\section{Analysis of symmetries of the dorsal stream baseline connectome}
\label{sec:comparison_connectome}

We perform a full symmetry analysis (including group and fibration
symmetry) of the dorsal stream baseline connectome in
\SuppFig{comparison_connectome}.

The automorphisms of the connectome are calculated with the VF2 algorithm \cite{cordella2004sub} as implemented in NetworkX \cite{hagberg2008exploring} (version 3.7). The connectome contains 11! =
39,916,800 possible permutations of its 11 ROIs. From this, only a few
are permutation symmetries.  The symmetry group of this connectome is generated by two non-trivial permutations, $\pi_1$ and $\pi_2$, and contains four elements:

\begin{equation}
\mathrm{Aut}(G) = \{\pi_0, \pi_1, \pi_2, \pi_3 = \pi_1 \pi_2 \} ,
\end{equation}
where the identity and the two generators in cycle notation are
(\SuppFig{comparison_connectome}b):

   \begin{equation}\begin{aligned}
   \pi_0\, =& \,\mbox{Id} \\
   \pi_1 \, =&\, ({\rm PreMA}_L\,{\rm PreMA}_R) \, ({\rm BA}_L\,{\rm BA}_R) \, ({\rm AG}_L\,{\rm AG}_R) \\
   & \, ({\rm WA}_L\, {\rm WA}_R) \,
    ({\rm SMG}_L\, {\rm SMG}_R) \\
    \pi_2  \, =\, &({\rm WA}_L\, {\rm SMG}_L) \, ({\rm WA}_R\, {\rm SMG}_R) 
    \end{aligned}\end{equation}

The actions of this symmetry group generate five orbits, which is the
orbital color partition that we see in Fig.\ref{fig:broken}a and in
\SuppFig{comparison_connectome}a:

   \begin{equation}\begin{aligned}
    \mathcal S_1=& \{{\rm PreMA}_L,{\rm PreMA}_R \} \\
    \mathcal S_2=& \{{\rm BA}_L, {\rm BA}_R \} \\
    \mathcal S_3=& \{{\rm AG}_L,{\rm AG}_R \} \\
    \mathcal S_4=& \{ {\rm WA}_L , {\rm WA}_R, {\rm SMG}_L, {\rm SMG}_R\} \\
    \mathcal S_5=& \{  {\rm SMA} \}
   \end{aligned}\end{equation}

The fibration analysis is done by searching for the minimal balanced
coloring of the network using the refinement algorithm of Kamei and
Cock \cite{kamei2013computation} employed by Morone et al.  in
\cite{morone2020fibration}.  The minimal balanced coloring of the
graph corresponds to a balanced coloring with a minimal number of
colors. Each cluster of balanced coloring is a fiber.  The resulting
coloring is shown in \SuppFig{comparison_connectome}a, left. It
shows the existence of five fibers; each fiber is also an orbit.  The minimal partition obtained by the fibers is the same as the
minimal orbital partition. This implies that the group symmetry of
this graph is the same as the fibration symmetry of the graph.

The fibers are:

   \begin{equation}\begin{aligned}
    \mathrm{Fiber}_1=& \{ {\rm PreMA}_L,{\rm PreMA}_R \} \\
    \mathrm{Fiber}_2=& \{ {\rm BA}_L, {\rm BA}_R \} \\
    \mathrm{Fiber}_3=& \{ {\rm AG}_L,{\rm AG}_R \} \\
    \mathrm{Fiber}_4=& \{ {\rm WA}_L , {\rm WA}_R, {\rm SMG}_L, {\rm SMG}_R \} \\
    \mathrm{Fiber}_5=& \{  {\rm SMA} \}
   \end{aligned}\end{equation}

Generally, the orbital partition obtained from automorphisms does not
necessarily need to coincide with the balanced coloring obtained from
the fibration analysis. Fibers capture more symmetries than orbits. Thus, the number of fibers is always equal to or smaller than the number
of orbits. Moreover, an orbit is always part of a fiber, but a fiber
may not be part of an orbit.  When these two partitions are the
same, the two symmetries are the same, too, implying a high symmetry state.

The analysis of the input trees of this connectome is shown in \SuppFig{comparison_connectome}c for the main fiber of 4 ROIs,
$\mathrm{Fiber}_4$ and a representative bilateral fiber $\mathrm{Fiber}_2$. This shows the isomorphism between the input trees of ROIs
within a fiber. This analysis complements the fibration analysis of
balanced coloring. The same analysis is done below for the RS and
task-based inferred state-specific structural subgraph.

\section{Integer linear program for symmetry-driven inference of the structural
  network to satisfy cluster synchronization}
\label{sec:algorithm}  

The way the functioning of the brain is connected to its underlying
structure adjusts according to the requirements of the brain activity
\cite{park2013structural,friston2011functional}.  Thus, the same
baseline highway can give rise to different functional states, e.g.,
RS or language task of verbal and fluency, given by different
synchronized coloring patterns (e.g., Fig.\ref{fig:functional}a, c,
e, respectively).  Consistent with this idea, a given functional
network (at rest or task) is sustained by a specific configuration of
the connectome, in a way, strictly depends on the activity itself.  A
condition for such matching to exist is that a modification of the
connectome displayed in Fig. \ref{fig:intro}a is introduced. 
Given the baseline connectome, which represents the communication
highway, only a subset of routes are needed to
guarantee the existence of the synchronous clusters (fibers/orbits) of
the functional network.

The crucial step of this scheme to infer the structural network from
the functional network is the fibration/group symmetry partitioning
and the iterative decimation process for the coloring matching. The partition problem is weakly NP-hard (NP, nondeterministic polynomial time) \cite{garey1978strong}, and it has
been shown that solving the directed/undirected in-balanced K-coloring
problem solves the partition problem \cite{leifer2022symmetry}, so that the coloring problem is NP-hard as well. In
particular, the problem of finding removals that satisfy the coloring
condition can be formulated as a mixed integer program that is
solvable for modestly sized instances. We follow a similar approach in this paper and formulate
the problem of finding the minimum perturbations to induce a minimal balanced coloring as a mixed integer linear program (MILP), an optimization problem with linear objective and constraint functions in which, in the present formulation, all decision variables are binary. We then solve the integer linear programs with the HiGHS solver \cite{huangfu2018parallelizing} through SciPy \cite{virtanen2020scipy} (version 1.18). In the analyses reported here only link removals are allowed ($a_{ij}=0$), and the minimality of the balanced coloring is verified on each solution by color refinement.

We consider a directed graph, $G = (V,E)$, where $V$ denotes the set of
nodes, and $E$ denotes the set of directed edges (an undirected edge is
considered as two directed edges). Also, we denote $n=|V|$ and $m=|E|$
as the number of nodes and directed edges, respectively. We also
define
\begin{equation}
E^C = \{ ij : i,j \in V, ij \not\in E\}
\end{equation}
 as the set of ordered
pairs of nodes for which no directed edge exists in $G$, which we refer to
as non-edges. These ordered
pairs represent potential edges that could be added to the graph
$G$. We let $\cS$ denote a coloring of $G$, i.e., $\cS$ is the
collection of sets partitioning $V$.  
This coloring, $\cS$, is provided by the CS from fMRI
synchronization in different engagements of the brain function,
Fig.\ref{fig:functional}a, c, e. We define $\alpha,
\beta$ as constant parameters that govern the objective's relative importance
between edge removal and edge addition.
We wish to determine the minimum number of edges to add or remove so that $\cS$ is a balanced coloring of $G$ (Methods Sect.~\ref{sec:fibration}). Our integer programs are guaranteed to find a balanced coloring but are not guaranteed to find a minimal balanced coloring. However, in our experiments, a minimal balanced coloring was found in all cases we tested.

The model's three families of binary decision variables are defined as follows. \\
For $ij \in E,$

\begin{equation}
    \label{eq:remove_vars}
r_{ij} = 1 \mbox{ if edge $ij$ is removed}, 0 \mbox{ otherwise.}
\end{equation}
For $ij \in E^C,$
\begin{equation}
    \label{eq:add_vars}
a_{ij} = 1 \mbox{ if non-edge $ij$ is added}, 0 \mbox{ otherwise.}
\end{equation}
For $P, Q, R \in \cS$ with $P\not=Q$ and for $i \in P, j \in Q$
\begin{equation}
    \label{eq:imbalance_vars}
s_{ijR} = 1 \mbox{ if $i$ and $j$ are imbalanced on $R$}
\end{equation}
and $0$ otherwise. The role of the linear constraints below are to set up a set of linear equalities and inequalities that, if satisfied by these decision variables, make $\cS$ a balanced coloring of the perturbed graph and favor its minimality (the minimality constraints below are necessary but not sufficient).\\

The objective function is to minimize the weighted sum of edges
removed and edges added.
The function is then defined as:
\begin{equation}
\label{eq:mip_obj}
f_{\alpha,\beta}(r,a) = \alpha \sum_{ij \in E} r_{ij} + \beta \sum_{ij \in E^C} a_{ij}.
\end{equation}

The main constraint assures that $\cS$ is a balanced coloring of the
perturbed graph $G$. 
\begin{align}
    \nonumber
    \sum_{ip \in E: i \in S} (1 - r_{ip}) + \sum_{ip \in E^C:i \in S} a_{ip} & = \\
    \sum_{iq \in E: i \in S} (1 - r_{iq}) + \sum_{iq \in E^C: i \in S} a_{iq}; 
   &p,q \in T; S,T \in \cS.
       \label{eq:balancing}
\end{align}

Constraints \eqref{eq:balancing} exist for every pair of nodes $p,q$
that are the same color and for every color set. Note that for a given
edge $ij \in E$, the quantity $1-r_{ij}$ is 1 if the edge is not
removed and 0 if it is removed. Also, for $ij \in E^C$, the quantity
$a_{ij}$ is 1 if $ij$ is a newly created edge and 0 otherwise. Thus,
the left-hand side of \eqref{eq:balancing} represents the edges that
enter into a given node $p$ from the color set $S$, and the right-hand
side represents the edges entering node $q$ from the color set
$S$. Using the same sums, \eqref{eq:atleastone} ensure that the
in-degree is at least one for every node:

\begin{equation}
    \label{eq:atleastone}
    \sum_{ip \in E} (1 - r_{ip}) + \sum_{ip} a_{ip}
    \geq 1, \quad \quad \quad p \in V.
\end{equation}

The following constraints are valid for minimal balanced colorings,
i.e., they are necessary but not sufficient.

\begin{align}
\nonumber
\sum_{ip \in E: i \in R} (1 - r_{ip}) + \sum_{ip:i \in R} a_{ip} - & \\
\nonumber
    \left(\sum_{iq \in E: i \in R} (1 - r_{iq}) + \sum_{iq: i \in R} a_{iq}\right)
    & \geq s_{pqR} - ns_{qpR};\\
\nonumber
& \\
p\in S; q\in T; R,S \not= T \in \cS,      \label{eq:minimal_nec1} \\
\nonumber
\sum_{iq \in E: i \in R} (1 - r_{iq}) + \sum_{iq:i \in R} a_{iq} - & \\
    \left(\sum_{ip \in E: i \in R} (1 - r_{ip}) + \sum_{ip: i \in R} a_{ip}\right)
    & \geq  s_{qpR} - ns_{pqR}; \nonumber \\
    p \in S; q\in T; R,S \not=T \in \cS,      \label{eq:minimal_nec2} \\
    \nonumber
    s_{pqR} + s_{qpR} &\leq 1; \\
    p\in S; q\in T; R,S \not= T \in \cS,      \label{eq:minimal_nec3} \\
    \sum_{R \in \cS} (s_{pqR} + s_{qpR})&  \geq 1;& \\
    p\in S; q\in T; S,T \in \cS 
    \label{eq:minimal_nec4} 
\end{align} 
The inequalities \eqref{eq:minimal_nec3} keep at most one of the two
binary variables $s_{pqR}$ to be equal to one for every color $R$. If
both are zero, then the inequalities \eqref{eq:minimal_nec1} and
\eqref{eq:minimal_nec2} would force $p$ and $q$ to be balanced for the
color $R$. If exactly one is equal to one, the numbers of in-neighbors of color $R$
of $p$ and $q$ differ by at least one. In particular, 
if $s_{pqR}=1$ and $s_{qpR}=0$, then the number of
in-neighbors of color $R$ of $p$ is at least one greater than that of
$q$. The converse is also true.
The inequalities \eqref{eq:minimal_nec4} force that one of $s_{pqR}$ or $s_{qpR}$ is equal to one for at least one color $R$. This is a necessary but not sufficient condition for the coloring to be minimal. For example, if two color partitions have no edges between them, the same number of edges to all other colors, and the same positive number of internal edges, then \eqref{eq:minimal_nec4} is satisfied as their different colors will register as an imbalance. However, the union of these two color partitions is balanced and has one less color, i.e., the coloring is no longer minimal. That being said, our experiments yielded strong evidence that the necessary condition sufficiently enforces the minimal balanced condition in practice, as we found a minimal balanced coloring for all of our test cases.

The complete model is then:
\begin{equation}
\begin{array}{ll}
\min f_{\alpha,\beta}(r,a) & \\[1mm]
\mbox{subject to} & \eqref{eq:balancing},
\eqref{eq:atleastone},
\eqref{eq:minimal_nec1}, \eqref{eq:minimal_nec2}, \eqref{eq:minimal_nec3},
\eqref{eq:minimal_nec4}, \\[1mm]
& r_{ij}, a_{k\ell}, s_{pqR} \in \{0,1\}, ij
\in E,\\[1mm]
& k\ell \in E^C, p \in P, q \in Q, P\not=Q,R \in \cS.
\end{array}
\label{eq:complete}
\end{equation}

The MILP operates on the binary connectome and returns the configurations with the minimum number of removed links (10 for the resting state and 14 for the tasks). Several configurations reach this minimum (16 for the resting state and 32 for the tasks, by exhaustive enumeration with the deletion-only formulation); all of them have the target fiber partition and no non-trivial automorphisms, so the symmetry analysis does not depend on which one is chosen. To select a single network we use the structural weights of the tracts as a percolation order: among the minimal configurations, we choose the one whose removed links are the weakest, comparing the removed tract weights from the strongest one downwards, with weights taken from the probabilistic diffusion-tractography connectome of Rosen and Halgren \cite{rosen2021whole} (Methods Sect.~\ref{sec:weighted}). This criterion uses structural information only and selects a unique network for each state (Fig.~\ref{fig:broken}b,c). In the resting state it retains the link from left BA to SMA, which is 50\% stronger than its right-hemisphere homolog and becomes part of the dominant SMA-BA$_L$ cluster during the tasks. Some of the remaining choices involve weight differences below the typical uncertainty of tractography (10--20\%) or callosal links, whose tractography weights are underestimated for lateral cortical regions because of crossing white-matter fibers \cite{jarbo2012vivo,ruddy2017transcallosal}; these choices are therefore the least constrained by the data. In the task network the callosal input WA$_R$~$\to$~WA$_L$ is present in every minimal configuration, since WA$_R$ is the only neighbour of WA$_L$ within its fiber, so its low tractography weight does not affect the selection.

Due to its large complexity, the brain can never have exact
symmetries, even within a single connectome. Structural brains are all
different, but a certain level of ideal symmetry must be common to all
of them to guarantee the performance of an average
synchronization pattern, despite not all structural
brains being identical. We apply the inference algorithms to those
group-average synchronization networks and connectomes at the
mesoscopic level, as shown in Fig.\ref{fig:functional} and
Fig.\ref{fig:broken} which should be interpreted as idealized networks.

\section{Symmetry breaking in physics and the brain}
\label{sec:breaking}

Most symmetry laws in physics are broken in one way or another. One
such mechanism is spontaneous symmetry breaking, where the laws of
physics remain symmetric, but the system's ground state exhibits a
lower symmetry than the full system, as in a
paramagnetic-to-ferromagnetic phase transition
\cite{yeomans1992statistical}. For temperatures below the critical
value $T_{\mathrm{C}}$, the magnetic moments of the atoms of ferromagnetic
material are partially aligned within magnetic domains, producing a net
magnetic moment even though the atoms interact through a spin-spin
interaction, which is invariant under rotation. Thus, the rotational
invariant symmetry of the system is broken into this ground state with
a non-zero magnetic moment. As the temperature increases, this
alignment is destroyed by thermal fluctuations and the net
magnetization progressively reduces until vanishing at $T_{\mathrm{C}}$. The
orientation of the magnetization is random. Each possible direction is
equally likely to occur, but only one is chosen at
random, resulting in a zero net magnetic moment. So, the rotational
symmetry of the ferromagnet is manifest for $T > T_{C}$ with zero
magnetic moment, but is broken by the arbitrary selection of a
particular (less-symmetrical) ground state with non-zero magnetic
moment for $T < T_{C}$.

Another type is explicit symmetry breaking, where the dynamics are
only approximately symmetric, yet the deviation caused by the breaking
forces is minimal. Hence, one can consider the symmetry violation as a
small correction in the system. An example is the spectral line
splitting in the Zeeman effect due to a magnetic interaction
perturbation in the Hamiltonian of the atoms involved.

In the present work, we implemented a symmetry-driven algorithm based
on a mixed integer linear program to infer the structural network
associated with each balanced coloring of the functional network
obtained experimentally in different tasks.  By applying this framework to healthy subjects performing standard language tasks, we
obtained a functional language anatomy which is consistent with the
common understanding of speech processing.

The symmetry-breaking we find in the brain is manifested in the
following:

\begin{enumerate}
\item The evidence of an underlying highly symmetrical connectome
  between the language areas (group symmetry = fibration symmetry) with a central fiber made of four ROIs. 
\item
  The evidence of a symmetrical language network representation during
  resting state as a consequence of the overall synchronization
  dynamics with a pentagonal fiber at the core of the network made of SMA, PreMA and WA. This network presents only fibration symmetries but no
  group symmetries thus, the resting state engagement breaks the
  global group symmetries of the baseline connectome. That is, even though the ROIs are synchronized in pairs with left-right symmetry, the underlying structural network does not have the global left-right symmetry. In fact, it has no non-trivial automorphisms, and the only surviving symmetry is the local fibration.
\item The characterization of the transition between resting state and
  language tasks as further broken symmetry, but this time of the
  fibration symmetry (the group symmetry remains broken). The evidence of a breaking of symmetry resulting in two central fibers (BA L-SMA) and (WA L-WA R-BA R).
  \item The evidence that the letter fluency and verb generation tasks, which share the same CS partition, are supported by the same pattern of communication routes, i.e., the same structural connectivity.
\item Possible applications will include the analysis of broken
  symmetries in neurological disorders and correlation with patients'
  clinical performance. Some neurological conditions compromise the
  synchronization in the brain (tumors, stroke, any focal lesion),
  affecting its coherent activity. By applying our method to these
  patients, we could shed light on bio-markers that could predict
  symptoms and patients' prognosis.

\end{enumerate}

We also note a limitation shared with the binary structural connectome itself: the present framework does not distinguish excitatory from inhibitory (sign-of-interaction) effects, and cluster-synchronization predictions under mixed-sign coupling can differ qualitatively from the all-positive case considered here. Extending the dynamical admissibility conditions of the fibration framework to signed and weighted interactions is an important direction for future work.

\section{Analysis of symmetries of the RS structural network}
\label{sec:comparison_rs}

We perform a full symmetry analysis (group and fibration symmetry) of
the inferred resting state structural network in \SuppFig{comparison_rs}.

First, the VF2 algorithm \cite{cordella2004sub} is used to calculate
the automorphisms of the network.
Out of the 11! =
39,916,800 possible permutations of its 11 ROIs only the
identity $[\pi_0\, = \,\mbox{Id}]$ is an automorphism
(\SuppFig{comparison_rs}b, left). Any other permutation is not a symmetry.
For instance, if we implement the permutation
\begin{equation}
\pi_2 \, =\,({\rm
  WA}_L\, {\rm SMG}_L) \, ({\rm WA}_R\, {\rm SMG}_R) ,
\end{equation}
we obtain a different graph (\SuppFig{comparison_rs}b,
right).

Accordingly, the resting state structural network has only trivial
group symmetry. Colloquially, we say that this network has no group
symmetry.  Since the only automorphism is the trivial identity, each node has its orbit. Thus, there are eleven orbits: one for
each node, \SuppFig{comparison_rs}a, right. If compared to the
baseline connectome that shows five orbits,
\SuppFig{comparison_connectome}a right, this represents a group
symmetry breaking.

The fibration symmetry analysis is done by finding the balanced
coloring (fibers) using the refinement algorithm of Kamei and Cock
\cite{kamei2013computation} and Morone et al.
\cite{morone2020fibration}.  \SuppFig{comparison_rs}a left shows the resulting minimal balanced coloring.  When compared to the baseline
connectome \SuppFig{comparison_connectome}a left, an increase of
fibration symmetry is obtained since now we observed a smaller number
of fibers. Recall that the most symmetric graph is that with a single
color, and the least symmetric graph is the one with $N$ colors for a
graph with $N$ nodes. While in the baseline, we have five fibers, in
the resting state the number of fibers is four.

When compared to the orbital partition of the same graph,
\SuppFig{comparison_rs}a, right, we find that this graph has
fibration symmetry but no group symmetry. The global symmetry has been
fully broken by engaging the brain in RS, but the local symmetry
remains, and it is enhanced in RS in comparison to the original
symmetry of the connectome.

The fibers are:

   \begin{equation}\begin{aligned}
    \mathrm{Fiber}_1=& \{ \rm SMA,{\rm PreMA}_L,{\rm PreMA}_R,{\rm WA}_L,{\rm WA}_R \} \\
    \mathrm{Fiber}_2=& \{ {\rm BA}_L, {\rm BA}_R \} \\
    \mathrm{Fiber}_3=& \{ {\rm AG}_L,{\rm AG}_R \} \\
    \mathrm{Fiber}_4=& \{ {\rm SMG}_L, {\rm SMG}_R \}
   \end{aligned}\end{equation}

The analysis of the input trees is shown in \SuppFig{comparison_rs}c for the main fiber of 5 ROIs, $\mathrm{Fiber}_1$ and a representative bilateral fiber $\mathrm{Fiber}_3$. This
complements the fibration analysis of this graph.

\section{Analysis of symmetries of the task structural network}
\label{sec:comparison_task}

We perform a full symmetry analysis (including group and fibration
symmetry) of the inferred task structural network in
\SuppFig{comparison_task}.

Similar to the RS network, the VF2 algorithm
\cite{cordella2004sub} shows that this graph has no automorphisms
except for the trivial identity.  From the 11! allowed permutations,
only the identity $[ \pi_0\, = \,\mbox{Id} ]$ is a symmetry
(\SuppFig{comparison_task}b, left). For instance, if we implement
the permutation
\begin{equation}
\pi_1 \, =\, ({\rm PreMA}_L\,{\rm
    PreMA}_R) \, ({\rm BA}_L\,{\rm BA}_R) \, ({\rm AG}_L\,{\rm AG}_R)
\, ({\rm WA}_L\, {\rm WA}_R) \, ({\rm SMG}_L\, {\rm SMG}_R)
\end{equation}
we obtain a different graph (\SuppFig{comparison_task}b, right),
exactly as we found for in RS (\SuppFig{comparison_rs}b).

Since there are no (non-trivial) automorphisms, we obtain eleven
orbits (one for each node, \SuppFig{comparison_task}a, right) for
the task network as in RS. Thus, the baseline connectome's global symmetry remains broken in the language task.

The fibration symmetry analysis for this connectome gives rise to a
the balanced coloring partition seen in \SuppFig{comparison_task}a, left.
We found five fibers:

   \begin{equation}\begin{aligned}
    \mathrm{Fiber}_1=& \{ \rm SMA,{\rm BA}_L \} \\
    \mathrm{Fiber}_2=& \{ {\rm PreMA}_L,{\rm PreMA}_R \} \\
    \mathrm{Fiber}_3=& \{ {\rm AG}_L,{\rm AG}_R \} \\
    \mathrm{Fiber}_4=& \{ {\rm BA}_R,{\rm WA}_L, {\rm WA}_R \} \\
    \mathrm{Fiber}_5=& \{ {\rm SMG}_L, {\rm SMG}_R \}
   \end{aligned}\end{equation}

If compared to the resting state connectome \SuppFig{comparison_rs}a, a decrease of fibration symmetry is
obtained. In RS, we have four fibers, and in the task, we have five. This
represents a fibration symmetry breaking (more fibers means less
fibration symmetry).  This local symmetry breaking is the product of
the breaking of left-right local symmetry in the Broca area due to language lateralization. Recall that the left-right global symmetry
has been fully broken in the RS state and remains broken
here. This local breaking of symmetry is done by the synchronization of
Broca left with SMA, and the independent synchronization of Broca
right to Wernicke left and right. These two areas remain locally
left-right symmetric. This produces the main two fibers controlling the
language network $\mathrm{Fiber}_1$ and $\mathrm{Fiber}_4$. The
analysis of the input trees of these fibers is shown in \SuppFig{comparison_task}c.

\section*{Data availability}
The raw magnetic resonance imaging (MRI) data acquired in this study at Memorial Sloan Kettering Cancer Center cannot be shared, in accordance with MSKCC policy: following the Health Insurance Portability and Accountability Act (HIPAA) and institutional regulations, all radiology imaging scans sent to other institutions for research or educational purposes must go through a de-identification process to ensure that protected health information (PHI) is protected before leaving the institution. The de-identified BOLD time series of the 11 ROIs for all participants and conditions, which are sufficient to reproduce all the analyses and figures of this paper, are available in the Open Science Framework (OSF) repository \url{https://osf.io/4ern8/} and in the code repository (\url{https://github.com/makselab/synchro-language-network}). The group-averaged PLV matrices can be regenerated from these time series with the code of the repository and are also provided in the Source Data file. The diffusion-tractography connectome used in this study is available from Rosen and Halgren \cite{rosen2021whole}. The group-consensus structural connectome used for comparison is available in the netneurolab/bazinet\_assortativity GitHub repository (\url{https://github.com/netneurolab/bazinet_assortativity}). The HCP-MMP1.0 (Glasser) parcellation is publicly available \cite{glasser2016multi}. Source data are provided with this paper.

\section*{Code availability}
The custom code developed in this study (symmetry analysis, mixed integer linear programming inference, cluster synchronization and whole-brain control analyses), together with the de-identified ROI time series needed to reproduce all analyses and figures, is available at \url{https://github.com/makselab/synchro-language-network} and \url{https://osf.io/4ern8/}, and is archived on Zenodo \cite{code_zenodo} (\url{https://doi.org/10.5281/zenodo.24094720}). The code is released under the MIT license and the data in the repository under the Creative Commons Attribution 4.0 International (CC BY 4.0) license. The analysis code was optimized with the assistance of the large language model Claude Opus 5.5 (Anthropic); the authors reviewed and tested all code and take full responsibility for it and for the results.

\clearpage

\begin{description}
 \item[Acknowledgements] We thank G. Del Ferraro and L. Alvarez for discussions.
 \item[Funding] This research was funded through the NIH U54 MSKCC-CCNY Partnership for Cancer Research. This work was also supported by the National Institute of Biomedical Imaging and Bioengineering (NIBIB) and the National Institute of Mental Health (NIMH) through the National Institutes of Health (NIH) BRAIN Initiative Grant R01 EB028157 (to M.Z. and H.A.M.) and by the Office of Naval Research (ONR) grant N00014-24-1-2758 (to H.A.M.). P.B. was partially funded by the project SERICS (PE00000014) under the National Recovery and Resilience Plan (NRRP) MUR programme funded by the EU NextGenerationEU. T.G. acknowledges financial support by the European Union NextGenerationEU, National Recovery and Resilience Plan (Piano Nazionale di Ripresa e Resilienza, PNRR), project SoBigData.it, Strengthening the Italian RI for Social Mining and Big Data Analytics, Grant IR0000013 (n. 3264, 28/12/2021). G.C. acknowledges funding from the Italian Ministero dell'Universit\`a e della Ricerca under the programme PRIN 2022 (re-ranking of the final lists), project number 2022KWTEB7, CUP B53C24006470006, and from the COGNIA visiting scheme at the London Institute for Mathematical Sciences (LIMS). The funders had no role in study design, data collection and analysis, decision to publish or preparation of the manuscript.
 \item[Author contributions] T.G. and H.A.M. conceived and designed the study. L.P. and A.H. acquired the fMRI data and defined the baseline connectome. T.G. performed the fMRI preprocessing and synchronization analyses. B.A., D.P. and P.B. developed the symmetry-inference algorithms. A.G., G.C. and M.Z. contributed to the interpretation of the results. T.G. and H.A.M. wrote the manuscript with input from all authors.
 \item[Competing interests] The authors declare no competing interests.
 \item[Correspondence] Correspondence and requests for materials should be addressed to H.A.M. \newline
 (email: hmakse@ccny.cuny.edu).
\end{description}

\clearpage
\begin{table}
\centering
\caption{{\bf Whole-brain control for shared input: resting state.} Cluster PLV (surrogate-corrected PLV averaged within each CS clique; $n=20$ participants) of the resting-state CS partition for raw data and for the nine-condition whole-brain regression battery (Methods Sect.~\ref{sec:wholebrain}). PLV, phase locking value; CS, cluster synchronization; SMA, supplementary motor area; BA, Broca's area; SMG, supramarginal gyrus; PreMA, premotor area; AG, angular gyrus; WA, Wernicke's area; L, left hemisphere; X-X, left-right homotopic pair of region X; A1, primary auditory cortex; PSL, perisylvian language area; MGN, medial geniculate nucleus. Raw, unregressed time series; global signal, mean signal over all 360 cortical parcels; global+A1+PSL and all combined, simultaneous regression of the indicated signals (all combined: global, A1, PSL, thalamus, MGN and inferior colliculus). \ding{51}, the clique satisfies Eq.~(\ref{eq:1}) in that condition, within the PLV tolerance of 0.01 of the control analyses (Methods Sect.~\ref{sec:wholebrain}). The clique with the lowest Cluster PLV in a condition is compared with no remaining ROI and satisfies Eq.~(\ref{eq:1}) by construction.}
\label{tab:wholebrain-rs}
\small
\begin{tabular}{lcccc}
\toprule
Condition & SMG-SMG & SMA-PreMA-PreMA-WA-WA & AG-AG & BA-BA \\
\midrule
raw                 & 0.746~\ding{51} & 0.697~\ding{51} & 0.689~\ding{51} & 0.645~\ding{51} \\
global signal       & 0.545~\ding{51} & 0.250~\ding{51} & 0.457~\ding{51} & 0.348~\ding{51} \\
A1                  & 0.526~\ding{51} & 0.324~\ding{51} & 0.595~\ding{51} & 0.442~\ding{51} \\
PSL                 & 0.338~\ding{51} & 0.440~\ding{51} & 0.510~\ding{51} & 0.429~\ding{51} \\
thalamus            & 0.707~\ding{51} & 0.628~\ding{51} & 0.610~\ding{51} & 0.580~\ding{51} \\
MGN                 & 0.672~\ding{51} & 0.572~\ding{51} & 0.572~\ding{51} & 0.550~\ding{51} \\
inferior colliculus & 0.689~\ding{51} & 0.614~\ding{51} & 0.591~\ding{51} & 0.584~\ding{51} \\
global+A1+PSL       & 0.286~\ding{51} & 0.202~\ding{51} & 0.405~\ding{51} & 0.296~\ding{51} \\
all combined        & 0.234~\ding{51} & 0.195~\ding{51} & 0.405~\ding{51} & 0.289~\ding{51} \\
\bottomrule
\end{tabular}
\end{table}

\begin{table}
\centering
\caption{{\bf Whole-brain control for shared input: letter fluency task.} Cluster PLV (surrogate-corrected PLV averaged within each CS clique; $n=20$ participants) of the task CS partition for raw task-state data and for the nine-condition whole-brain regression battery (Methods Sect.~\ref{sec:task-generalization}). PLV, phase locking value; CS, cluster synchronization; SMA, supplementary motor area; BA, Broca's area; SMG, supramarginal gyrus; PreMA, premotor area; AG, angular gyrus; WA, Wernicke's area; L, left hemisphere; X-X, left-right homotopic pair of region X; A1, primary auditory cortex; PSL, perisylvian language area; MGN, medial geniculate nucleus. Raw, unregressed time series; global signal, mean signal over all 360 cortical parcels; global+A1+PSL and all combined, simultaneous regression of the indicated signals (all combined: global, A1, PSL, thalamus, MGN and inferior colliculus). \ding{51}, the clique satisfies Eq.~(\ref{eq:1}) in that condition, within the PLV tolerance of 0.01 of the control analyses (Methods Sect.~\ref{sec:wholebrain}). The clique with the lowest Cluster PLV in a condition is compared with no remaining ROI and satisfies Eq.~(\ref{eq:1}) by construction.}
\label{tab:suppl-lf}
\small
\begin{tabular}{lccccc}
\toprule
Condition & SMA-BA$_{\mathrm{L}}$ & PreMA-PreMA & AG-AG & BA$_{\mathrm{R}}$-WA-WA & SMG-SMG \\
\midrule
raw                 & 0.789~\ding{51} & 0.747~\ding{51} & 0.667~\ding{51} & 0.624~\ding{51} & 0.659~\ding{51} \\
global signal       & 0.728~\ding{51} & 0.572~\ding{51} & 0.538~\ding{51} & 0.334~\ding{51} & 0.498~\ding{51} \\
A1                  & 0.781~\ding{51} & 0.671~\ding{51} & 0.572~\ding{51} & 0.455~\ding{51} & 0.459~\ding{51} \\
PSL                 & 0.775~\ding{51} & 0.754~\ding{51} & 0.526~\ding{51} & 0.487~\ding{51} & 0.370~\ding{51} \\
thalamus            & 0.700~\ding{51} & 0.608~\ding{51} & 0.624~\ding{51} & 0.456~\ding{51} & 0.627~\ding{51} \\
MGN                 & 0.743~\ding{51} & 0.625~\ding{51} & 0.581~\ding{51} & 0.494~\ding{51} & 0.600~\ding{51} \\
inferior colliculus & 0.775~\ding{51} & 0.735~\ding{51} & 0.664~\ding{51} & 0.578~\ding{51} & 0.653~\ding{51} \\
global+A1+PSL       & 0.612~\ding{51} & 0.577~\ding{51} & 0.492~\ding{51} & 0.313~\ding{51} & 0.276~\ding{51} \\
all combined        & 0.552~\ding{51} & 0.568~\ding{51} & 0.475~\ding{51} & 0.293~\ding{51} & 0.271~\ding{51} \\
\bottomrule
\end{tabular}
\end{table}

\begin{table}
\centering
\caption{{\bf Whole-brain control for shared input: verb generation task.} Cluster PLV (surrogate-corrected PLV averaged within each CS clique; $n=20$ participants) of the task CS partition for raw task-state data and for the nine-condition whole-brain regression battery (Methods Sect.~\ref{sec:task-generalization}). The SMA-BA$_{\mathrm{L}}$ cluster has Cluster PLV $=0.714$ in the main analysis (Fig.~\ref{fig:functional}e,f) and $0.827$ in the whole-brain pipeline used here, in which time series are averaged within Glasser parcels rather than extracted from 6 mm spheres. PLV, phase locking value; CS, cluster synchronization; SMA, supplementary motor area; BA, Broca's area; SMG, supramarginal gyrus; PreMA, premotor area; AG, angular gyrus; WA, Wernicke's area; L, left hemisphere; X-X, left-right homotopic pair of region X; A1, primary auditory cortex; PSL, perisylvian language area; MGN, medial geniculate nucleus. Raw, unregressed time series; global signal, mean signal over all 360 cortical parcels; global+A1+PSL and all combined, simultaneous regression of the indicated signals (all combined: global, A1, PSL, thalamus, MGN and inferior colliculus). \ding{51}, the clique satisfies Eq.~(\ref{eq:1}) in that condition, within the PLV tolerance of 0.01 of the control analyses (Methods Sect.~\ref{sec:wholebrain}). The clique with the lowest Cluster PLV in a condition is compared with no remaining ROI and satisfies Eq.~(\ref{eq:1}) by construction.}
\label{tab:suppl-vg}
\small
\begin{tabular}{lccccc}
\toprule
Condition & SMA-BA$_{\mathrm{L}}$ & PreMA-PreMA & AG-AG & BA$_{\mathrm{R}}$-WA-WA & SMG-SMG \\
\midrule
raw                 & 0.827~\ding{51} & 0.688~\ding{51} & 0.581~\ding{51} & 0.606~\ding{51} & 0.597~\ding{51} \\
global signal       & 0.796~\ding{51} & 0.399~\ding{51} & 0.398~\ding{51} & 0.302~\ding{51} & 0.449~\ding{51} \\
A1                  & 0.776~\ding{51} & 0.557~\ding{51} & 0.494~\ding{51} & 0.436~\ding{51} & 0.487~\ding{51} \\
PSL                 & 0.823~\ding{51} & 0.625~\ding{51} & 0.422~\ding{51} & 0.448~\ding{51} & 0.303~\ding{51} \\
thalamus            & 0.689~\ding{51} & 0.543~\ding{51} & 0.431~\ding{51} & 0.444~\ding{51} & 0.517~\ding{51} \\
MGN                 & 0.751~\ding{51} & 0.529~\ding{51} & 0.479~\ding{51} & 0.432~\ding{51} & 0.539~\ding{51} \\
inferior colliculus & 0.753~\ding{51} & 0.612~\ding{51} & 0.546~\ding{51} & 0.552~\ding{51} & 0.587~\ding{51} \\
global+A1+PSL       & 0.665~\ding{51} & 0.397~\ding{51} & 0.424~\ding{51} & 0.254~\ding{51} & 0.293~\ding{51} \\
all combined        & 0.608~\ding{51} & 0.484~\ding{51} & 0.386~\ding{51} & 0.242~\ding{51} & 0.272~\ding{51} \\
\bottomrule
\end{tabular}
\end{table}

\clearpage

\clearpage

\end{document}


\begin{center}
{\Large\bf Supplementary Information}\\[4mm]
{\large Fibration symmetry-breaking supports functional transitions in a brain network engaged in language}\\[3mm]
Tommaso Gili, Bryant Avila, Luca Pasquini, Andrei Holodny, David Phillips, Paolo Boldi, Andrea Gabrielli, Guido Caldarelli, Manuel Zimmer \& Hern\'an A. Makse
\end{center}

\vspace{4mm}
\noindent Supplementary Figures 1--13 are numbered in the order in which they are first cited in the main text. Unless stated otherwise, group-level results refer to $n=20$ participants. Cluster synchronization (CS) cliques are defined by Eq.~(1) of the main text; PLV denotes the phase locking value.

\clearpage

\begin{figure}[p]
\centering
\includegraphics[width=\textwidth]{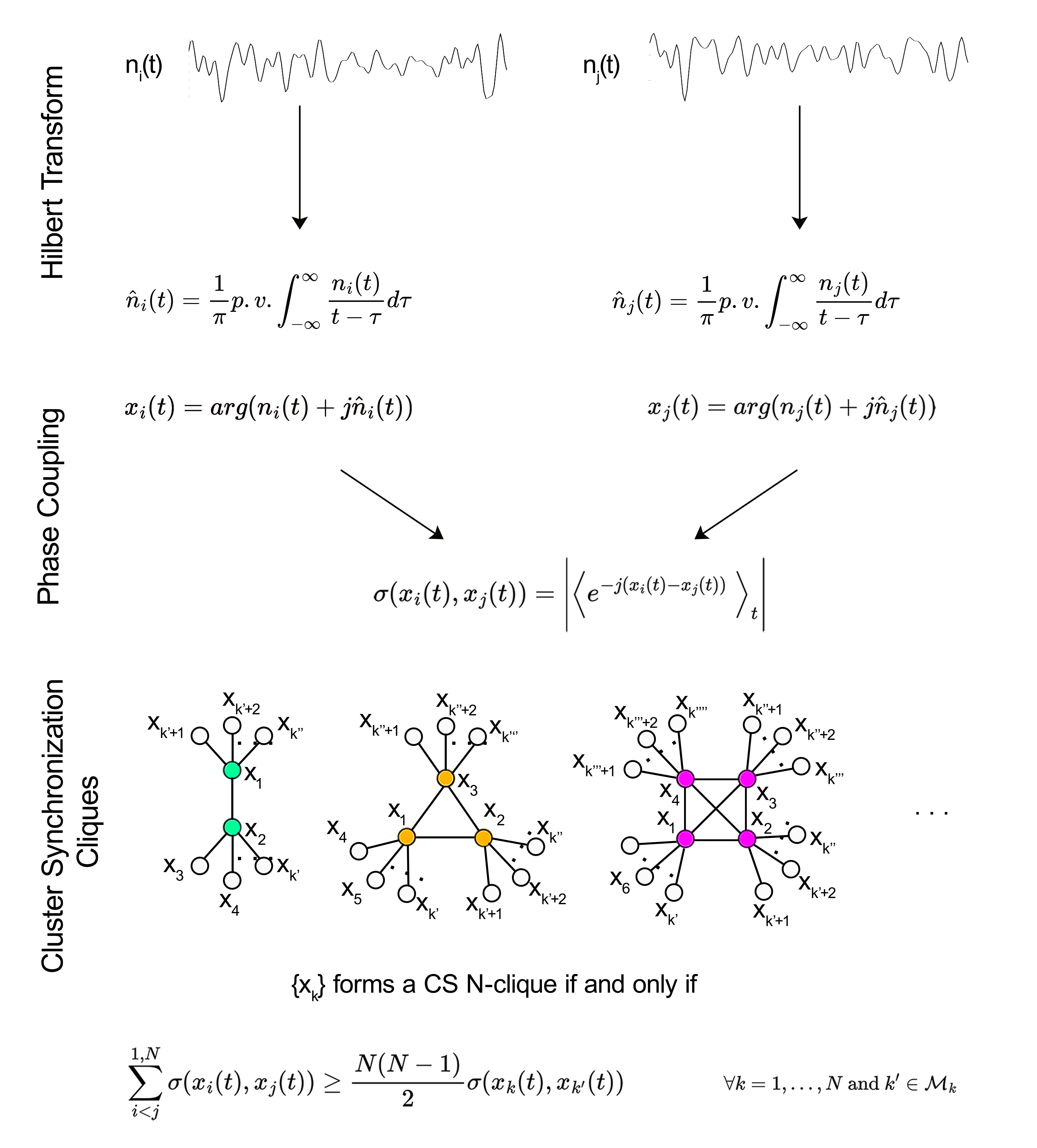}
\caption{{\bf Schematic of the synchronization clustering algorithm.} Pairs of blood oxygen level-dependent (BOLD) time series from pairs of cerebral regions of interest (ROIs) are Hilbert transformed to obtain their instantaneous phases, which enter the phase locking value (PLV) calculation. Once all pairs of ROIs have been included, the synchronization clustering algorithm is applied to the resulting PLV matrix to identify the cluster synchronization (CS) cliques.}
\label{fig:plv}
\end{figure}

\begin{figure}[p]
\centering
\includegraphics[width=\textwidth,height=0.78\textheight,keepaspectratio]{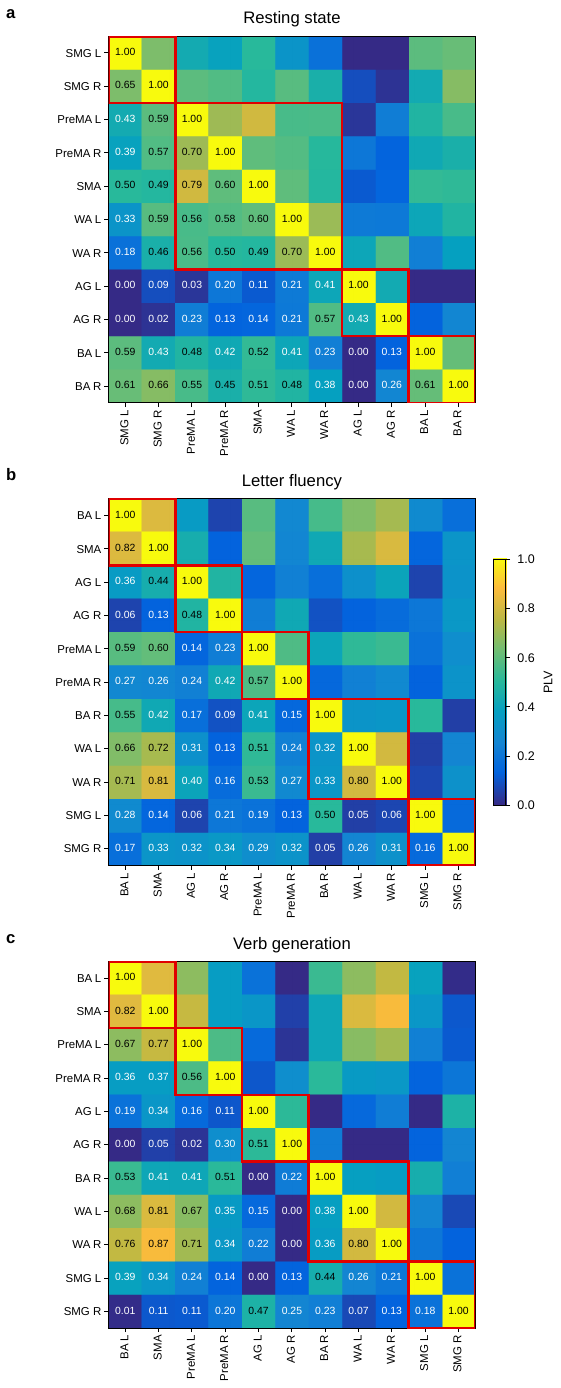}
\caption{{\bf Single-participant phase locking value matrices.} Phase locking value (PLV) corrected against surrogate data (Eq.~(2) of the main text Methods) for participant P13, whose resting-state matrix is the most similar to the group average. {\bf a} Resting state. {\bf b} Letter fluency task. {\bf c} Verb generation task. Red boxes indicate the group-level cluster synchronization (CS) cliques of Fig.~2 of the main text, shown in the same order. Numbers are PLV values; color scale from 0 (no phase locking) to 1 (complete phase locking). \roiabbr}
\label{fig:correlation}
\end{figure}

\begin{figure}[p]
\centering
\includegraphics[width=\textwidth]{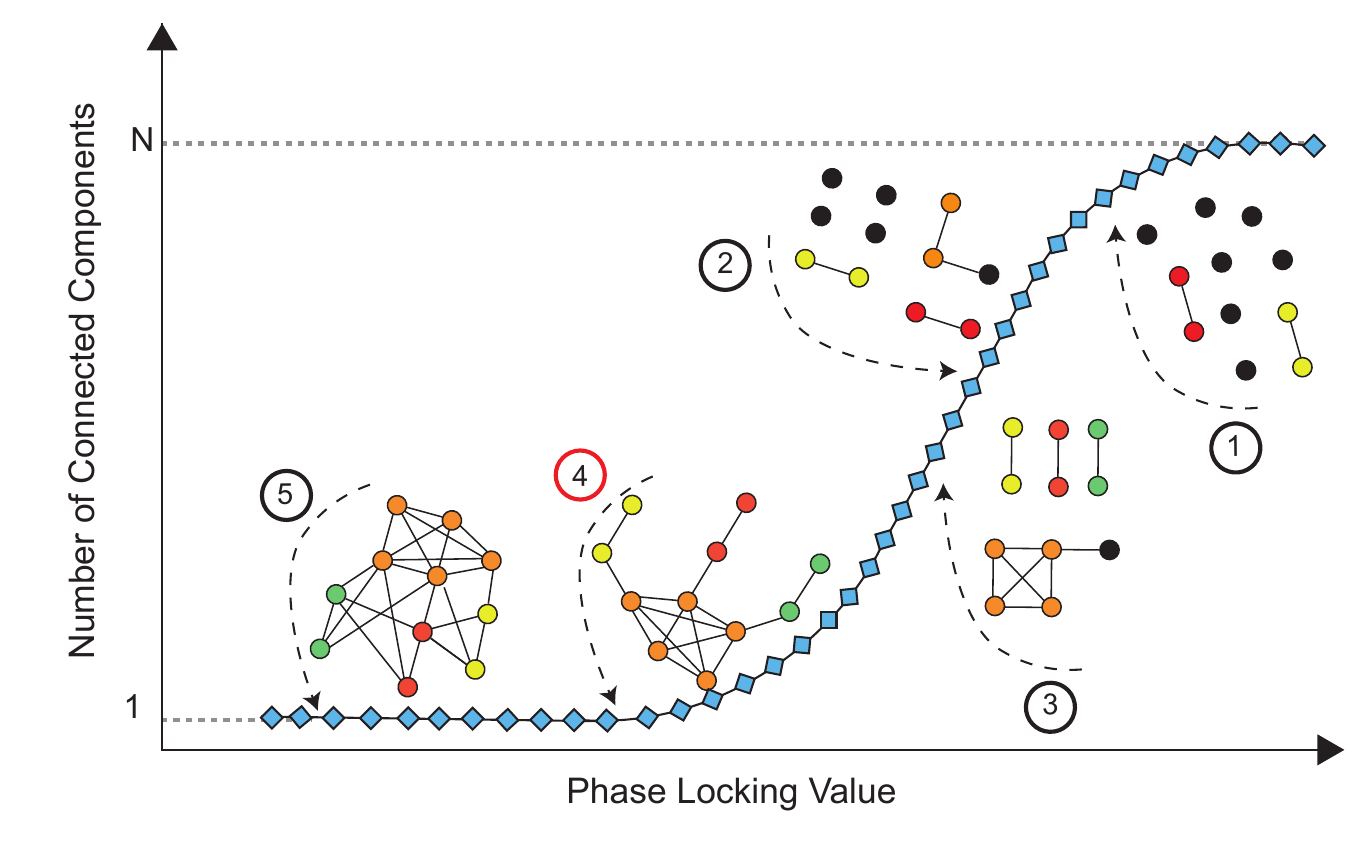}
\caption{{\bf Percolation procedure for the definition of the cluster synchronization network in a toy model.} Candidate clusters of synchronized nodes are identified by a percolation procedure on the synchronization matrix: starting from a disconnected graph, links are added in decreasing order of weight (synchronization), starting from the largest. Candidate cliques are accepted as cluster synchronization (CS) cliques when, ordered by decreasing Cluster PLV, each satisfies Eq.~(1) of the main text. Blue diamonds show the number of connected components as a function of the phase locking value (PLV). Nodes, initially black, are colored as soon as they are included in a clique. Five PLV thresholds are shown, corresponding to five connectivity patterns of the toy model; the fourth corresponds to the PLV below which the process stops.}
\label{fig:perolation_cliques}
\end{figure}

\begin{figure}[p]
\centering
\includegraphics[width=\textwidth,height=0.773\textheight,keepaspectratio]{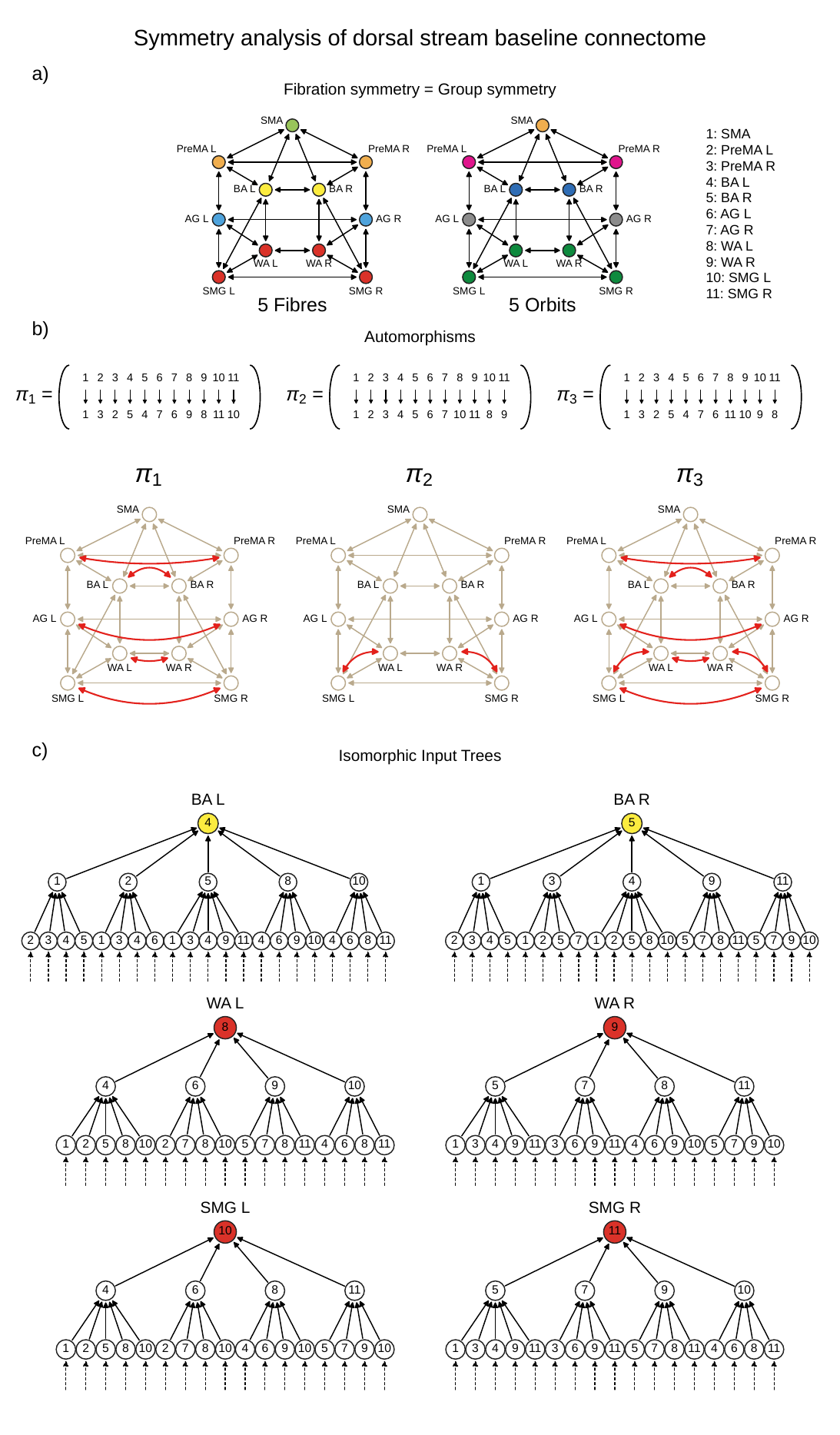}
\caption{{\bf Group and fibration symmetry analysis of the dual stream (dorsal) baseline connectome of language.} {\bf a} The orbital partition (automorphisms) and the fiber partition (fibrations) of this network coincide, indicating a high level of symmetry of the baseline connectome. {\bf b} Generators of the symmetry group of the baseline connectome. Left: the left-right (mirror) global symmetry $\pi_1$. Center: the symmetry permutation $\pi_2 = (\mathrm{WA_L}\,\mathrm{SMG_L})\,(\mathrm{WA_R}\,\mathrm{SMG_R})$. Right: the composition $\pi_3 = \pi_1 \pi_2$. Applying these symmetries to each node generates the orbits. {\bf c} Isomorphic input trees of the main fiber made of four ROIs (WA and SMG, left and right) and of a representative bilateral fiber (BA, left and right); the remaining bilateral fibers are analogous. Colors denote fibers (balanced coloring). \roiabbr}
\label{fig:comparison_connectome}
\end{figure}

\begin{figure}[p]
\centering
\includegraphics[width=\textwidth,height=0.733\textheight,keepaspectratio]{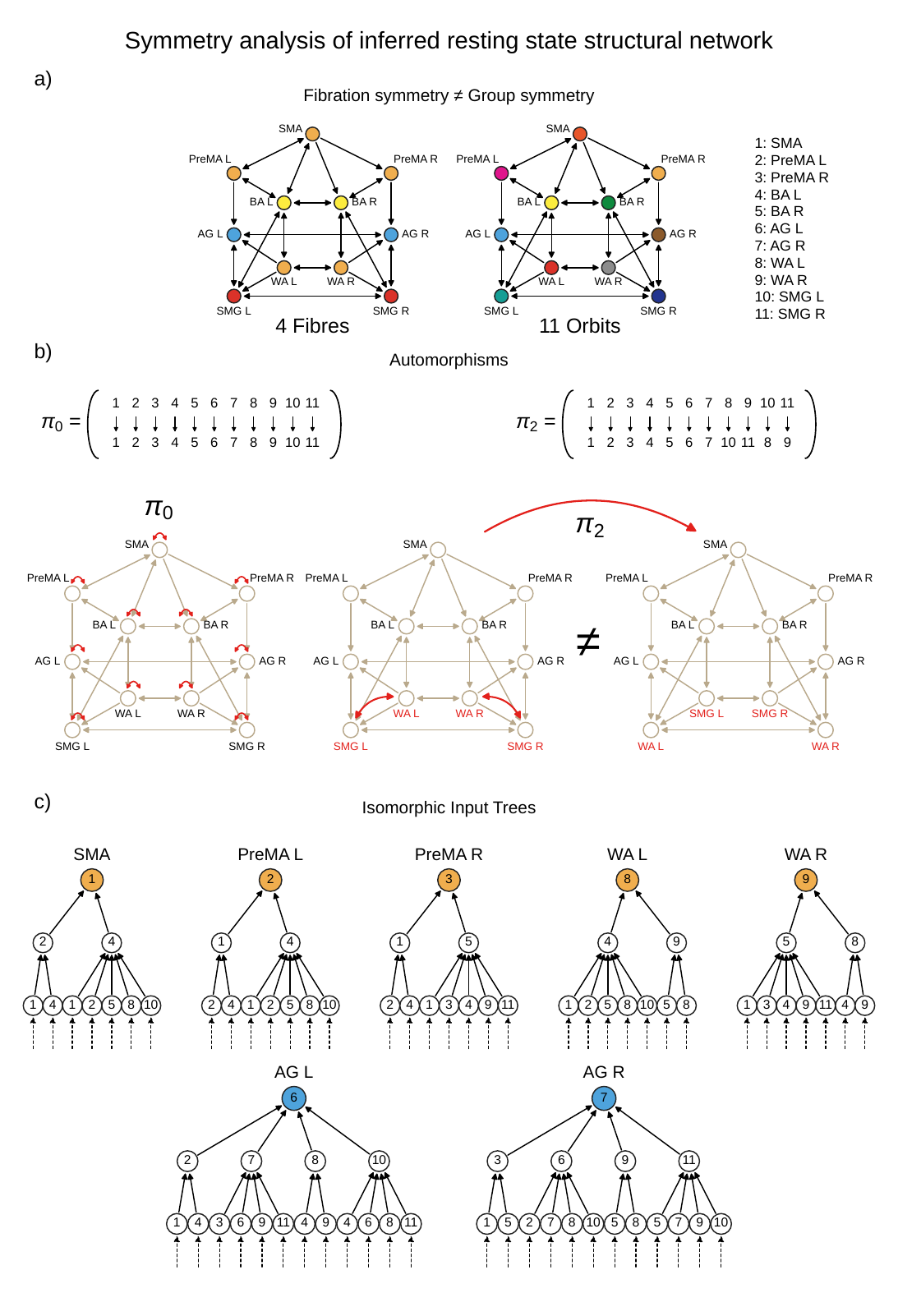}
\caption{{\bf Group and fibration symmetry analysis of the inferred structural network supporting the resting state.} {\bf a} Fiber partition (left) and orbital partition (right) of the inferred resting-state structural network. The network has no automorphisms except the identity, which gives a trivial orbital partition of 11 colors in which each ROI is its own orbit. The symmetry group of the baseline connectome (Supplementary Fig.~\ref{fig:comparison_connectome}a) is therefore broken in the resting state, whereas fibration symmetry remains: the fibration analysis identifies four fibers, shown as four balanced colors. {\bf b} Only the identity $\pi_0$ is a global symmetry (left); the permutation $\pi_2$ (right) is not a symmetry, although WA left and right remain locally symmetric under a fibration. {\bf c} Isomorphic input trees of the ROIs in the main, pentagonal fiber (PreMA left and right, SMA, WA left and right) and of a representative bilateral fiber (AG left and right); the remaining bilateral fibers are analogous. \roiabbr}
\label{fig:comparison_rs}
\end{figure}

\begin{figure}[p]
\centering
\includegraphics[width=\textwidth,height=0.733\textheight,keepaspectratio]{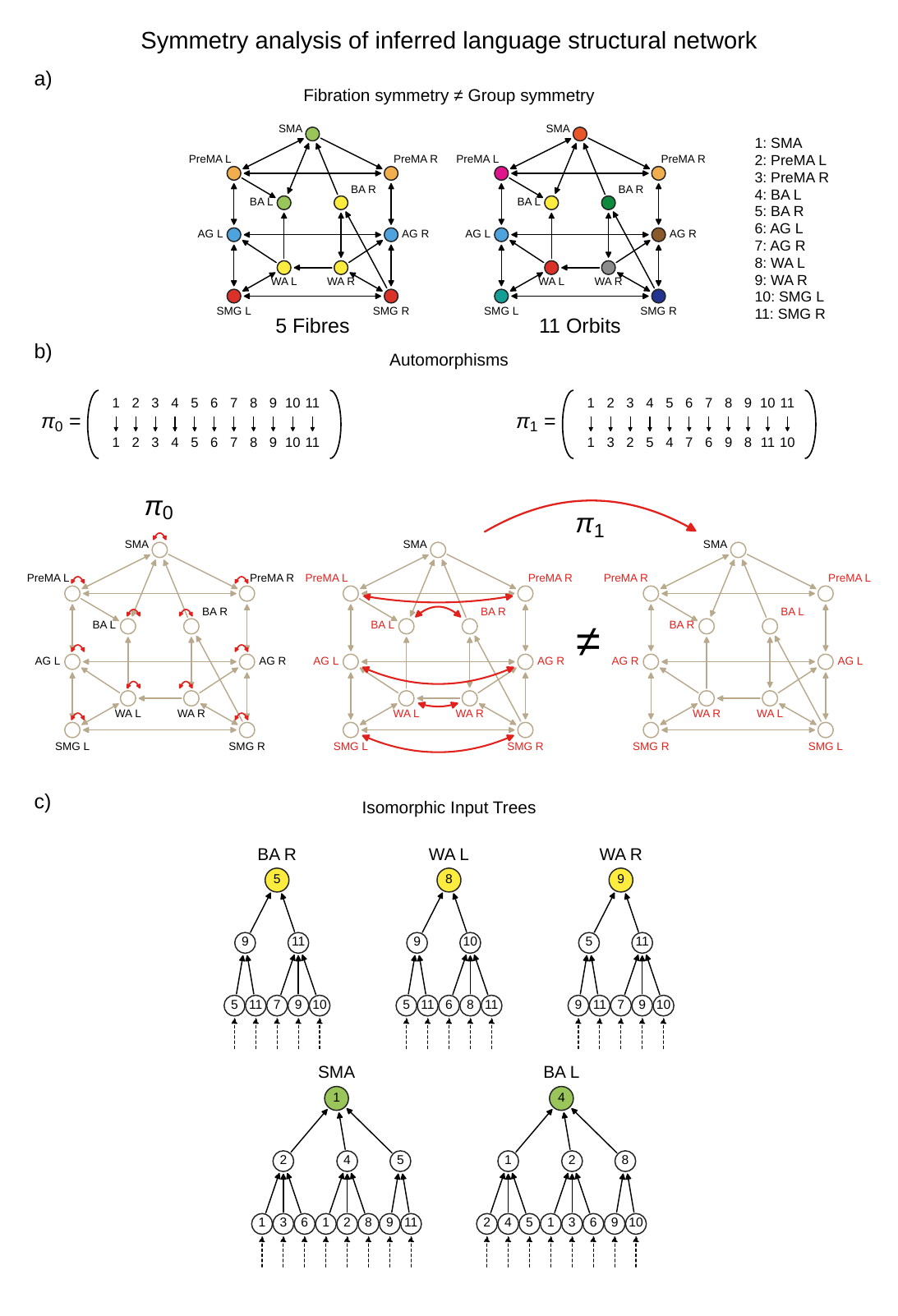}
\caption{{\bf Group and fibration symmetry analysis of the inferred structural network supporting the language task.} {\bf a} Fiber partition (left) and orbital partition (right) of the inferred task structural network. The network has no automorphisms except the identity, which gives a trivial orbital partition of 11 colors. The symmetry group remains broken in the task, as in the resting state (Supplementary Fig.~\ref{fig:comparison_rs}a). The fibration symmetry of the resting state is broken by task-induced lateralization: BA left and BA right are recruited by SMA and WA, respectively, and belong to two different fibers. The number of fibers increases to five, indicating lower fibration symmetry than in the resting state. {\bf b} As in the resting state, the network has no non-trivial automorphisms; for example, the global left-right permutation $\pi_1$ is not a symmetry. {\bf c} Isomorphic input trees of the ROIs in the largest fiber (WA left and right, BA right) and in the fiber formed by SMA and BA left. \roiabbr}
\label{fig:comparison_task}
\end{figure}

\begin{figure}[p]
\centering
\includegraphics[width=0.8\textwidth]{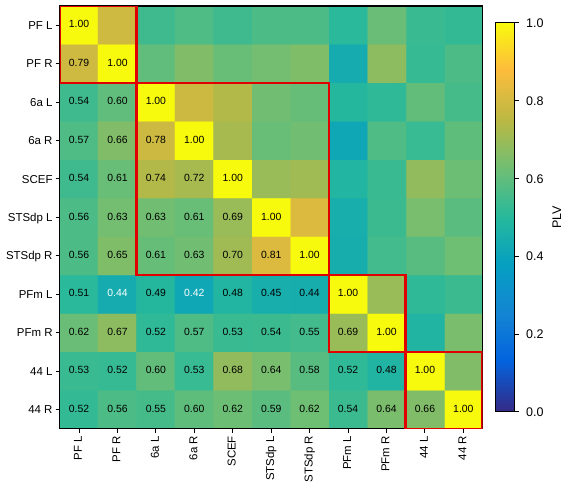}
\caption{{\bf Resting-state PLV matrix from time series averaged within regions of the Glasser atlas.} Group-averaged phase locking value (PLV) corrected against surrogate data (Eq.~(2) of the main text Methods) matrix ($n=20$ participants) for the resting state, computed from time series averaged within the Glasser parcels of the dorsal stream of the language system: PF, SMG; 6a, PreMA; SCEF, SMA; STSdp, WA; PFm, AG; 44, BA. The left and right SCEF parcels were merged into a single ROI, consistent with the ROI definition in the main text. Red boxes indicate the cluster synchronization (CS) cliques, which satisfy Eq.~(1) of the main text with the same partition found with spherical ROIs. Numbers are PLV values. \roiabbr}
\label{fig:extended_figure_glasser}
\end{figure}

\begin{figure}[p]
\centering
\includegraphics[width=\textwidth]{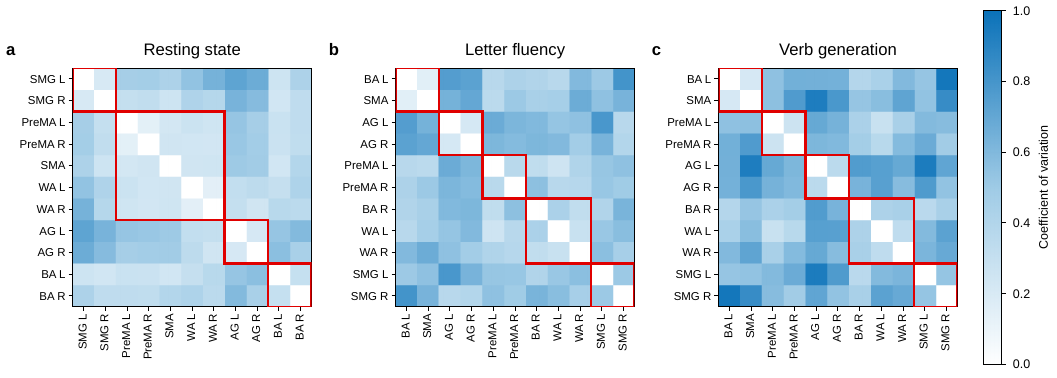}
\caption{{\bf Coefficient of variation of the phase locking values across participants.} Coefficient of variation (standard deviation divided by the mean across $n=20$ participants) of the phase locking value (PLV) corrected against surrogate data (Eq.~(2) of the main text Methods) for each pair of ROIs. {\bf a} Resting state. {\bf b} Letter fluency task. {\bf c} Verb generation task. Red boxes indicate the cluster synchronization (CS) cliques; clusters are shown in the order used for the group-averaged matrices in Fig.~2 of the main text. \roiabbr}
\label{fig:variationcoefficient}
\end{figure}

\begin{figure}[p]
\centering
\includegraphics[width=\textwidth]{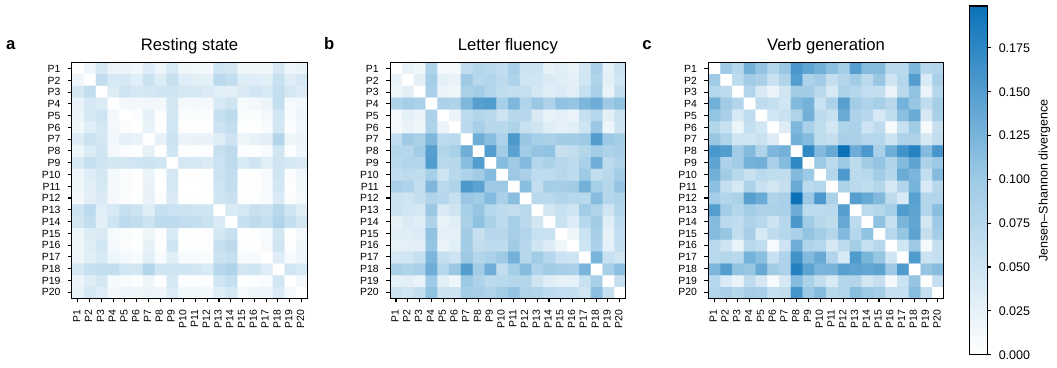}
\caption{{\bf Jensen-Shannon divergence between participants.} Jensen-Shannon divergence between the distributions of the PLV values of each pair of participants (P1--P20), where each participant's distribution is given by the phase locking value (PLV) corrected against surrogate data (Eq.~(2) of the main text Methods) of all ROI pairs, normalized to unit sum. {\bf a} Resting state. {\bf b} Letter fluency task. {\bf c} Verb generation task.}
\label{fig:jsd}
\end{figure}

\begin{figure}[p]
\centering
\includegraphics[width=0.8\textwidth]{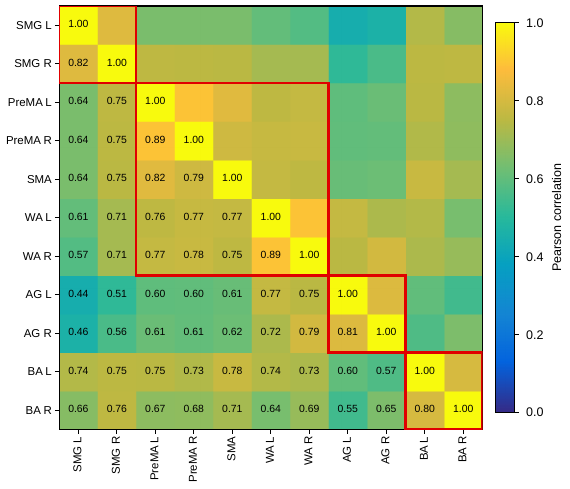}
\caption{{\bf Resting-state correlation matrix.} Group-averaged Pearson correlation matrix of the blood oxygen level-dependent (BOLD) time series of the 11 ROIs in the resting state ($n=20$ participants). Red boxes indicate the cluster synchronization (CS) cliques obtained from the phase locking value (Fig.~2b of the main text), which also satisfy Eq.~(1) of the main text when the correlation is used as the synchronization measure. Numbers are correlation values. \roiabbr}
\label{fig:extended_figure_12}
\end{figure}

\begin{figure}[p]
\centering
\includegraphics[width=0.8\textwidth]{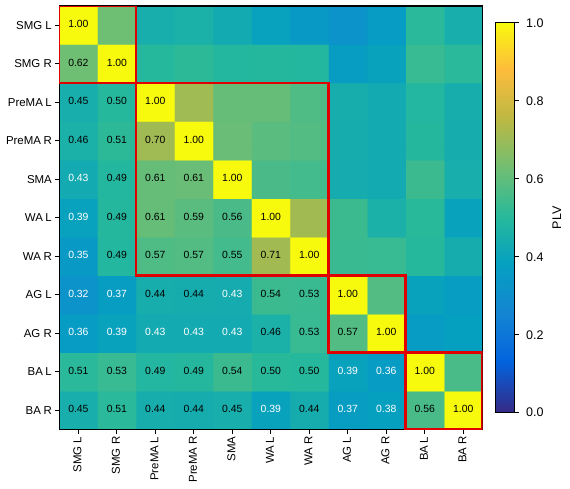}
\caption{{\bf Resting-state PLV matrix from time series deconvolved by the hemodynamic response function.} Group-averaged phase locking value (PLV) matrix, corrected against surrogate data (Eq.~(2) of the main text Methods), computed from resting-state time series after estimation and deconvolution of the hemodynamic response function (HRF) ($n=20$ participants). Red boxes indicate the cluster synchronization (CS) cliques of Fig.~2b of the main text, which also satisfy Eq.~(1) of the main text in the deconvolved data. Numbers are PLV values. \roiabbr}
\label{fig:extended_figure_deconvolved}
\end{figure}

\begin{figure}[p]
\centering
\includegraphics[width=0.8\textwidth]{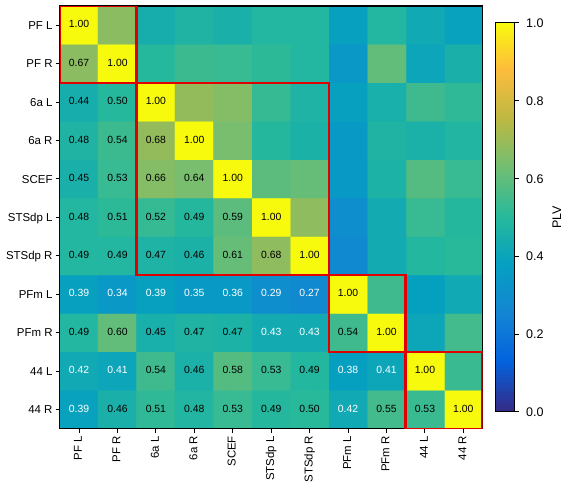}
\caption{{\bf Resting-state PLV matrix from HRF-deconvolved time series averaged within regions of the Glasser atlas.} Group-averaged phase locking value (PLV) matrix, corrected against surrogate data (Eq.~(2) of the main text Methods), computed from resting-state time series deconvolved by the hemodynamic response function (HRF) and averaged within the Glasser parcels of the dorsal stream of the language system ($n=20$ participants): PF, SMG; 6a, PreMA; SCEF, SMA; STSdp, WA; PFm, AG; 44, BA. The left and right SCEF parcels were merged into a single ROI. Red boxes indicate the cluster synchronization (CS) cliques of Fig.~2b of the main text. The four clusters satisfy Eq.~(1) of the main text within the PLV tolerance of 0.01 used for the control analyses (main text Methods). Numbers are PLV values. \roiabbr}
\label{fig:extended_figure_deconvolved_glasser}
\end{figure}

\begin{figure}[p]
\centering
\includegraphics[width=\textwidth]{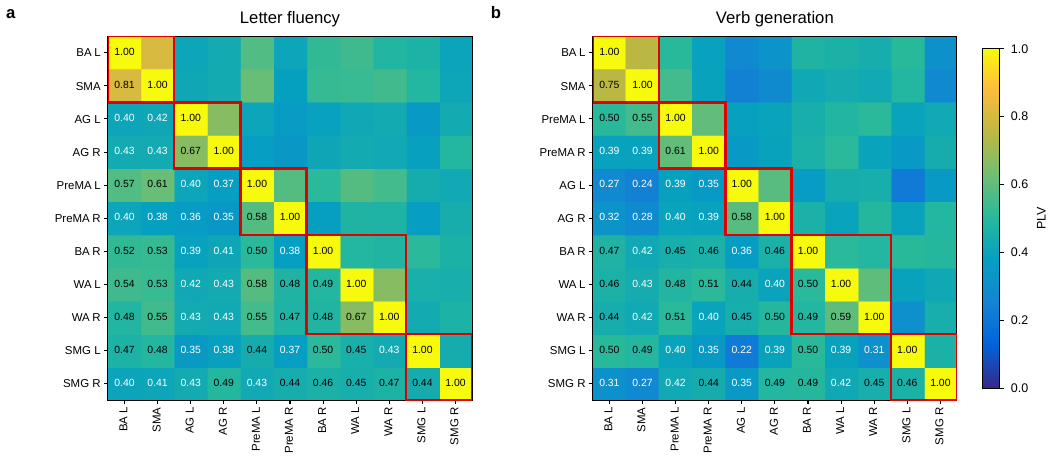}
\caption{{\bf Task PLV matrices from band-pass filtered time series.} Group-averaged phase locking value (PLV) matrices, corrected against surrogate data (Eq.~(2) of the main text Methods), computed from task time series band-pass filtered in the frequency range used for the resting state (0.01--0.1 Hz) ($n=20$ participants). {\bf a} Letter fluency task. {\bf b} Verb generation task. Numbers are PLV values. Red boxes indicate the cluster synchronization (CS) cliques of the main analysis (Fig.~2 of the main text), shown in the same order, which also satisfy Eq.~(1) of the main text in the band-pass filtered data. \roiabbr}
\label{fig:bandpass}
\end{figure}